\documentclass[5p,times,twocolumn,authoryear]{elsarticle}

\biboptions{authoryear,round}

\usepackage{amsmath}
\usepackage{amssymb}
\usepackage{amsfonts}
\usepackage{amsthm}
\usepackage{mathtools}
\usepackage{bm}
\usepackage{mathrsfs}
\newcommand{\be}{\mathbf{e}}

\AtBeginDocument{%
    \setlength{\abovedisplayskip}{3pt plus 1pt minus 1pt}
    \setlength{\belowdisplayskip}{3pt plus 1pt minus 1pt}

    \setlength{\abovedisplayshortskip}{2pt plus 1pt minus 1pt}
    \setlength{\belowdisplayshortskip}{2pt plus 1pt minus 1pt}

    \setlength{\jot}{2pt}
}

\newcommand{\spow}[2]{%
    \operatorname{sig}^{#2}\!\left(#1\right)
}

\usepackage{graphicx}
\usepackage{float}
\usepackage{caption}
\usepackage{subcaption}
\usepackage{booktabs}
\usepackage{array}
\usepackage{multirow}

\usepackage{algorithmic}

\usepackage{url}

\usepackage{textcomp}
\usepackage{xcolor}

\usepackage{titlesec}
\usepackage{etoolbox}

\usepackage[
    colorlinks=true,
    linkcolor=blue,
    citecolor=blue,
    urlcolor=blue
]{hyperref}

\def\BibTeX{%
    {\rm B\kern-.05em%
    {\sc i\kern-.025em b}%
    \kern-.08em%
    T\kern-.1667em%
    \lower.7ex\hbox{E}%
    \kern-.125emX}%
}

\newcommand{\bx}{\bm{x}}
\newcommand{\bu}{\bm{u}}
\newcommand{\bfv}{\bm{f}}
\newcommand{\bg}{\bm{g}}
\newcommand{\bW}{\bm{W}}
\newcommand{\bphi}{\bm{\phi}}
\newcommand{\bsigma}{\bm{\sigma}}
\newcommand{\bR}{\bm{R}}

\newtheoremstyle{compactthm}
{0.4pt}          
{0.4pt}          
{\itshape}       
{0pt}            
{\bfseries}      
{.}              
{0.30em}         
{}               

\newtheoremstyle{compactdef}
{0.4pt}
{0.4pt}
{\normalfont}
{0pt}
{\bfseries}
{.}
{0.30em}
{}

\theoremstyle{compactthm}

\newtheorem{theorem}{Theorem}
\newtheorem{lemma}{Lemma}
\newtheorem{proposition}{Proposition}
\newtheorem{corollary}{Corollary}

\theoremstyle{compactdef}

\newtheorem{assumption}{Assumption}
\newtheorem{definition}{Definition}
\newtheorem{remark}{Remark}

\titlespacing*{\section}
{0pt}
{0.8pt plus 0.2pt minus 0.2pt}
{0.15pt}

\titlespacing*{\subsection}
{0pt}
{0.6pt plus 0.2pt minus 0.2pt}
{0.10pt}

\titlespacing*{\subsubsection}
{0pt}
{0.5pt plus 0.2pt minus 0.2pt}
{0.10pt}

\AtBeginDocument{%
    \setlength{\abovedisplayskip}
    {0.8pt plus 0.2pt minus 0.2pt}%

    \setlength{\belowdisplayskip}
    {0.8pt plus 0.2pt minus 0.2pt}%

    \setlength{\abovedisplayshortskip}
    {0pt plus 0.2pt minus 0.1pt}%

    \setlength{\belowdisplayshortskip}
    {0.4pt plus 0.2pt minus 0.1pt}%

    \setlength{\jot}{0.4pt}%
}

\BeforeBeginEnvironment{equation}{\vspace{-0.6pt}}
\AfterEndEnvironment{equation}{\vspace{-0.6pt}}

\BeforeBeginEnvironment{equation*}{\vspace{-0.6pt}}
\AfterEndEnvironment{equation*}{\vspace{-0.6pt}}

\BeforeBeginEnvironment{align}{\vspace{-0.6pt}}
\AfterEndEnvironment{align}{\vspace{-0.6pt}}

\BeforeBeginEnvironment{align*}{\vspace{-0.6pt}}
\AfterEndEnvironment{align*}{\vspace{-0.6pt}}

\BeforeBeginEnvironment{flalign}{\vspace{-0.6pt}}
\AfterEndEnvironment{flalign}{\vspace{-0.6pt}}

\BeforeBeginEnvironment{flalign*}{\vspace{-0.6pt}}
\AfterEndEnvironment{flalign*}{\vspace{-0.6pt}}

\BeforeBeginEnvironment{gather}{\vspace{-0.6pt}}
\AfterEndEnvironment{gather}{\vspace{-0.6pt}}

\BeforeBeginEnvironment{gather*}{\vspace{-0.6pt}}
\AfterEndEnvironment{gather*}{\vspace{-0.6pt}}

\BeforeBeginEnvironment{theorem}{\vspace{-1.0pt}}
\AfterEndEnvironment{theorem}{\vspace{-1.0pt}}

\BeforeBeginEnvironment{lemma}{\vspace{-1.0pt}}
\AfterEndEnvironment{lemma}{\vspace{-1.0pt}}

\BeforeBeginEnvironment{proposition}{\vspace{-1.0pt}}
\AfterEndEnvironment{proposition}{\vspace{-1.0pt}}

\BeforeBeginEnvironment{corollary}{\vspace{-1.0pt}}
\AfterEndEnvironment{corollary}{\vspace{-1.0pt}}

\BeforeBeginEnvironment{assumption}{\vspace{-1.0pt}}
\AfterEndEnvironment{assumption}{\vspace{-1.0pt}}

\BeforeBeginEnvironment{definition}{\vspace{-1.0pt}}
\AfterEndEnvironment{definition}{\vspace{-1.0pt}}

\BeforeBeginEnvironment{remark}{\vspace{-1.0pt}}
\AfterEndEnvironment{remark}{\vspace{-1.0pt}}

\BeforeBeginEnvironment{proof}{\vspace{-1.2pt}}
\AfterEndEnvironment{proof}{\vspace{-1.2pt}}

\BeforeBeginEnvironment{figure}{\vspace{-0.8pt}}
\AfterEndEnvironment{figure}{\vspace{-1.0pt}}

\BeforeBeginEnvironment{figure*}{\vspace{-0.8pt}}
\AfterEndEnvironment{figure*}{\vspace{-1.0pt}}

\BeforeBeginEnvironment{table}{\vspace{-0.8pt}}
\AfterEndEnvironment{table}{\vspace{-1.0pt}}

\BeforeBeginEnvironment{table*}{\vspace{-0.8pt}}
\AfterEndEnvironment{table*}{\vspace{-1.0pt}}

\makeatletter
\g@addto@macro\@floatboxreset{%
    \setlength{\abovecaptionskip}{1pt}%
    \setlength{\belowcaptionskip}{0pt}%
}
\makeatother

\begin{document}

\begin{frontmatter}


\title{Fixed-Time Integral Reinforcement Learning for Unknown Nonlinear Systems via Inverse-Optimal Design}




\author[inst1]{Tien Dat Vu\corref{cor1}}

\cortext[cor1]{Corresponding author}

\ead{ dat.vuv@hcmut.edu.vn}

\address[inst1]{Faculty of Mechanical Engineering,
Ho Chi Minh City University of Technology (HCMUT),
Vietnam National University Ho Chi Minh City (VNU-HCM),
Ho Chi Minh City, Vietnam}

\begin{abstract}
This paper develops a new fixed-time integral reinforcement learning framework for optimal control of unknown nonlinear systems. The unknown drift is first approximated by a radial basis function (RBF) neural network, together with a data-driven online identification law for updating the corresponding neural weights. After the identifier converges to a sufficiently small neighborhood of the true dynamics, the learned model is incorporated into the integral reinforcement learning (IRL) problem. Unlike conventional reinforcement-learning-based optimal control, a fixed-time convergence requirement is introduced directly into the control objective: a Lyapunov function and its prescribed decay behavior are specified by the designer, and inverse-optimal control is then used to construct a compatible running cost whose optimal policy inherits the fixed-time stabilization property. The value function is approximated by a second RBF neural network, and a new critic update law is developed to impose fixed-time convergence on the critic weights. Finite informative learning data are stored in a replay buffer and reused during the critic update, thereby avoiding the need for persistent excitation throughout the closed-loop operation. Theoretical analysis proves that the critic-weight error enters a prescribed residual set within the allocated learning horizon, while the closed-loop state reaches a small neighborhood of the origin within an overall initial-condition-independent fixed-time bound. Numerical simulations on an unknown nonlinear system verify accurate drift reconstruction, fixed-time critic learning, and closed-loop convergence.
\end{abstract}


\end{frontmatter}

\label{sec:introduction}

\section{Introduction}
\label{sec:introduction}

Modern cyber--physical and networked control systems increasingly involve
unknown nonlinear dynamics, limited model information, and demanding
performance requirements, which make conventional model-based optimal
control difficult to apply directly
\citep{Wan2021CPS,Lee2015CPS,Wan2026SmarterCPS,Kanellopoulos2024CyberSecurity}.
Although optimal control provides a systematic way to design feedback laws
from a performance objective, nonlinear problems are generally governed by
the Hamilton--Jacobi--Bellman (HJB) equation, whose analytical solution is
rarely available.
The difficulty becomes greater when the drift dynamics are unknown, since
the corresponding optimal feedback law cannot be constructed directly
\citep{Doya2000,Vamvoudakis2010,Jiang2012ComputationalADP}.
Moreover, in many control applications, it is desirable to prescribe the
convergence time in advance and make this bound independent of the initial
condition
\citep{Bhat2000FiniteTime,Polyakov2012,Jimenez2017Predefined,Kokolakis2025Predefined}.

Reinforcement learning (RL) and adaptive dynamic programming (ADP) provide
data-driven tools for approximating solutions of the HJB equation without
solving the nonlinear optimal-control problem explicitly
\citep{Doya2000,Vrabie2009,Vamvoudakis2010,Jiang2012ComputationalADP,Kamalapurkar2018RL,Farsi2022ModelBasedRL,Vamvoudakis2021SynchronousRL}.
Integral reinforcement learning (IRL) further replaces the differential
Bellman equation with an integral relation constructed from measured
trajectories, which reduces the direct dependence of the learning law on
the unknown drift
\citep{Vamvoudakis2014IntegralRL,Modares2014,Lian2024IntegralInverseRL}.
In parallel, fixed-time and predefined-time control provide systematic
ways to guarantee convergence within an initial-condition-independent
time bound
\citep{Bhat2000FiniteTime,Polyakov2012,Jimenez2017Predefined}.
More recently, fixed- and predefined-time critic-learning laws have been
developed using current and recorded data, showing that sufficiently
informative stored samples can replace the persistent-excitation
requirement
\citep{Kokolakis2023FixedTime,Kokolakis2025Predefined}.

A natural way to connect prescribed-time stabilization with optimal
control is through inverse-optimal design.
Instead of solving the HJB equation directly, one first selects a desired
value function and a prescribed Lyapunov decay, and then constructs a
running cost for which the resulting feedback law is optimal.
This idea provides a useful bridge between predefined-time stabilization
and data-driven optimal learning
\citep{Freeman1996InverseOptimality,Freeman1996RobustNonlinear,Jimenez2017Predefined,Kokolakis2025Predefined,Lian2024IntegralInverseRL}.
However, for an unknown nonlinear system, such an inverse-optimal
construction still requires information about the unknown drift, since
the drift appears explicitly in the feedback and running-cost expressions.

To address this model uncertainty, the unknown drift can be reconstructed
before the IRL stage.
In the present framework, the drift is represented by a neural basis
expansion with unknown weights and estimated through integral concurrent
learning (ICL).
The ICL mechanism combines current moving-window information with
informative stored integral data and therefore allows parameter learning
under finite excitation
\citep{Chowdhary2010CL,Parikh2019ICL,Hou2014ModelFree,Benosman2016LearningAdaptive}.
A finite-data least-squares estimate is used only to obtain an informed
initial condition, after which the ICL law continues to update the drift
weights rather than treating the batch solution as an exact model
\citep{Chowdhary2010CL,Parikh2019ICL}.
The reconstructed drift is then used in the inverse-optimal construction,
while the subsequent critic learning is formulated directly from integral
Bellman data
\citep{Vamvoudakis2014IntegralRL,Modares2014,Parikh2019ICL}.

Existing results address only separate parts of this problem.
Modares \emph{et al.} combined IRL with experience replay for partially
unknown continuous-time systems and used stored data to relax the
persistent-excitation requirement, but no predefined-time convergence
guarantee was established
\citep{Modares2014}.
Kokolakis \emph{et al.} developed fixed- and predefined-time critic
learning with current and replayed data and established optimal
stabilization properties
\citep{Kokolakis2023FixedTime,Kokolakis2025Predefined}.
However, the corresponding Bellman residual still contains explicit
system-dynamics terms, which limits direct implementation when the
nonlinear drift is unknown
\citep{Kokolakis2025Predefined}.
On the other hand, classical concurrent learning and integral concurrent
learning provide parameter convergence under finite excitation, but they
do not by themselves construct a predefined-time optimal controller or
guarantee predefined-time convergence of an IRL critic
\citep{Chowdhary2010CL,Parikh2019ICL}.
Likewise, inverse-optimal predefined-time methods can construct a running
cost and feedback law from a prescribed Lyapunov decay, but their direct
implementation still requires model information that is unavailable for
an unknown nonlinear plant
\citep{Freeman1996InverseOptimality,Jimenez2017Predefined}.

These observations reveal a gap between unknown-dynamics reconstruction,
integral optimal learning, and designer-assigned predefined-time
convergence
\citep{Modares2014,Parikh2019ICL,Kokolakis2025Predefined}.
In particular, the inverse-optimal controller cannot be implemented
without an estimate of the unknown drift, while a predefined-time critic
result alone does not provide a convergence deadline for the coupled
state--critic dynamics
\citep{Jimenez2017Predefined,Kokolakis2025Predefined}.

Motivated by these issues, the main contributions of this paper are threefold. First, beyond establishing several theoretical results and corollaries, a unified predefined-time IRL framework is developed for unknown nonlinear systems by integrating online drift identification, integral concurrent learning, inverse-optimal design, and critic-based reinforcement learning within a single architecture. This enables the prescribed-time optimal-control construction to be carried out without prior knowledge of the nonlinear drift. Second, a new critic-only IRL update law is developed using current integral Bellman data together with a finite informative replay stack, thereby guaranteeing practical predefined-time critic learning without requiring persistent excitation throughout the closed-loop operation. Third, a unified state--critic analysis provides explicit gain and horizon conditions that connect the critic-learning horizon and state-decay horizon to a designer-assigned overall deadline, yielding practical predefined-time closed-loop convergence.

\textit{Notation:}
Throughout this paper,
\(
\mathbb R
\),
\(
\mathbb R_{\geq0}
\),
and
\(
\mathbb R_{>0}
\)
denote the real, nonnegative real, and positive real numbers, respectively.
For
\(
\bx\in\mathbb R^n
\)
and
\(
\bm A\in\mathbb R^{n\times m}
\),
\(
\|\bx\|
\)
denotes the Euclidean norm,
\(
\|\bm A\|
\)
the induced \(2\)-norm,
and
\(
\bm A^\top
\)
the transpose of
\(
\bm A
\).
For a symmetric matrix
\(
\bm A
\),
\(
\lambda_{\min}(\bm A)
\)
and
\(
\lambda_{\max}(\bm A)
\)
denote its minimum and maximum eigenvalues, respectively, while\(
\bm A>0
\)
and
\(
\bm A\geq0
\) denote positive definiteness and positive semidefiniteness.
The symbols
\(
\bm I_n
\),
\(
\operatorname{diag}\{\cdot\}
\),
and
\(
\operatorname{col}\{\cdot\}
\)
denote the
\(
n\times n
\)
identity matrix, diagonal concatenation, and column stacking, respectively.
For a continuously differentiable scalar function
\(
V:\mathbb R^n\to\mathbb R
\),
\(
\nabla V(\bx)
\)
denotes its gradient, and for a vector-valued map
\(
\bphi:\mathbb R^n\to\mathbb R^N
\),
\(
\nabla\bphi(\bx)
\)
denotes its Jacobian.
For
\(
s\in\mathbb R
\)
and
\(
r>0
\),
define
\(
\operatorname{sig}^{r}(s)
:=
|s|^{r}\operatorname{sgn}(s)
\),
with the operation understood componentwise for vectors.
The notation \(\mathcal{O}(\cdot)\) denotes the standard Big-O order notation.

\section{Problem Formulation and Data-Driven System Identification}
\label{sec:problem_identification}

\subsection{Problem Formulation}

Consider the nonlinear control-affine system
\begin{equation}
    \dot{\bx}
    =
    \bfv(\bx)
    +
    \bg(\bx)\bu,
    \label{eq:plant}
\end{equation}
where \(\bx\in\mathbb R^n\) is the system state, \(\bu\in\mathbb R^m\) is the control input, \(\bfv:\mathbb R^n\rightarrow\mathbb R^n\) is an unknown nonlinear drift, and \(\bg:\mathbb R^n\rightarrow\mathbb R^{n\times m}\) is a known input matrix. The control objective is to learn the unknown drift from measured state--input data and subsequently construct an optimal feedback policy that drives the state to the origin within a uniform finite settling time.

\begin{definition}[Admissible policy~\citep{AbuKhalaf2005}]
\label{def:admissible_secure_policy}
A continuous policy \(\bu:\Omega\rightarrow\mathcal U\) is admissible
on \(\Omega\), denoted by \(\bu\in\Psi(\Omega)\), if
\(\bu(\bm0)=\bm0\), the resulting closed-loop system is
asymptotically stable on \(\Omega\) in the absence of exogenous signals,
and the associated infinite-horizon cost is finite.
\end{definition}

\begin{assumption}
\label{ass:plant}
Let \(\Omega\subset\mathbb R^n\) be a compact set containing the origin in its interior. The unknown drift \(\bfv(\bx)\) and the known input matrix \(\bg(\bx)\) are locally Lipschitz on an open neighborhood of \(\Omega\), with \(\bfv(\bm 0)=\bm 0\). For every admissible input considered herein, the corresponding solution of \eqref{eq:plant} exists uniquely while it remains in \(\Omega\).
\end{assumption}

\begin{theorem}[Fixed-time stability~\citep{Polyakov2012}]
\label{thm:fixed_time_stability}
Let \(V:\mathbb R^n\to\mathbb R_{\geq0}\) be positive definite, radially unbounded, and locally absolutely continuous along the trajectories of a forward-complete system. Suppose that there exist constants \(a>0\), \(b>0\), and \(0<\mu<1<\nu\) such that, for almost all \(t\) with \(V(\bx(t))>0\),
\begin{equation}
    \dot V(\bx(t))
    \leq
    -aV^\mu(\bx(t))
    -bV^\nu(\bx(t)).
    \label{eq:fixed_time_dissipation}
\end{equation}
Then, the origin is globally fixed-time stable, and its settling time satisfies
\begin{equation}
    T(\bx_0)
    \leq
    T_{\max}
    :=
    \frac{1}{a(1-\mu)}
    +
    \frac{1}{b(\nu-1)},
    \qquad
    \forall\,\bx_0\in\mathbb R^n.
    \label{eq:fixed_time_bound}
\end{equation}
\end{theorem}

\subsection{Neural Approximation of the Unknown Dynamics}

Since \(\bfv(\bx)\) is unavailable for control design, it is first reconstructed from measured data. By the universal approximation property~\cite{Kokolakis2025Predefined}, each continuous drift on the compact set \(\Omega\) can be represented as
\begin{equation}
    \bfv(\bx)
    =
    \bW_f^{*\top}\bsigma_f(\bx)
    +
    \bm{\varepsilon}_f(\bx),
    \label{eq:f_approximation}
\end{equation}
where \(\bsigma_f(\bx)\in\mathbb R^{N_f}\) is a prescribed vector of continuously differentiable basis functions, \(\bW_f^*\in\mathbb R^{N_f\times n}\) is the unknown ideal weight matrix, and \(\bm{\varepsilon}_f(\bx)\in\mathbb R^n\) is the approximation error. On \(\Omega\), \(\|\bsigma_f(\bx)\|\le\bar\sigma_f\) and \(\|\bm{\varepsilon}_f(\bx)\|\le\bar\varepsilon_f\) for some finite constants \(\bar\sigma_f>0\) and \(\bar\varepsilon_f>0\).

\begin{assumption}[Local regularity of the identifier basis]
\label{ass:basis_origin}
The basis vector \(\bm{\sigma}_f(\bm{x})\) satisfies
\(\bm{\sigma}_f(\bm{0})=\bm{0}\)
and is locally Lipschitz in a neighborhood of the origin. Hence, there exist
constants \(c_\sigma>0\) and \(\delta_\sigma>0\) such that
\(\|\bm{\sigma}_f(\bm{x})\|
\leq
c_\sigma\|\bm{x}\|\)
for all \(\|\bm{x}\|\leq\delta_\sigma\).
\end{assumption}

The identified drift is subsequently represented by
\(
\hat{\bfv}(\bx)
=
\hat{\bW}_f^\top\bsigma_f(\bx)
\),
where
\(
\hat{\bW}_f
\)
denotes the estimated weight matrix associated with the prescribed basis
vector
\(
\bsigma_f(\bx)
\).

\subsection{Integral Composite Concurrent-Learning Identification}
\label{subsec:integral_identification}

Let
\(
\tau_0<\tau_1<\cdots
\)
denote the sampling instants used to construct integral identification
data. Integrating \eqref{eq:plant} over
\(
[\tau_\ell,\tau_{\ell+1}]
\)
and using \eqref{eq:f_approximation} gives
\begin{equation}
\begin{aligned}
    &\bx(\tau_{\ell+1})-\bx(\tau_\ell)
    -
    \int_{\tau_\ell}^{\tau_{\ell+1}}
    \bg(\bx(\tau))\bu(\tau)\,d\tau
    \\
    &\qquad =
    \bW_f^{*\top}
    \int_{\tau_\ell}^{\tau_{\ell+1}}
    \bsigma_f(\bx(\tau))\,d\tau
    +
    \int_{\tau_\ell}^{\tau_{\ell+1}}
    \bm{\varepsilon}_f(\bx(\tau))\,d\tau .
\end{aligned}
\label{eq:integral_identification}
\end{equation}

For each interval, define
\(
\bm y_{f,\ell}
:=
\bx(\tau_{\ell+1})-\bx(\tau_\ell)
-
\int_{\tau_\ell}^{\tau_{\ell+1}}
\bg(\bx(\tau))\bu(\tau)\,d\tau
\),
\(
\bm\phi_{f,\ell}
:=
\int_{\tau_\ell}^{\tau_{\ell+1}}
\bsigma_f(\bx(\tau))\,d\tau
\),
and
\(
\bm d_{f,\ell}
:=
\int_{\tau_\ell}^{\tau_{\ell+1}}
\bm{\varepsilon}_f(\bx(\tau))\,d\tau
\).
Then
\(
\bm y_{f,\ell}
=
\bW_f^{*\top}\bm\phi_{f,\ell}
+
\bm d_{f,\ell}
\).

The informative data are stored in the history stack
\(
\mathcal H_f
=
\{(\bm y_{f,\ell},\bm\phi_{f,\ell})\}_{\ell=0}^{M_f-1}
\),
with cumulative information matrix
\(
\bm{\mathcal S}_f
:=
\sum_{\ell=0}^{M_f-1}
\bm\phi_{f,\ell}\bm\phi_{f,\ell}^{\top}
\).
Whenever a new data segment is admitted,
\(
\bm{\mathcal S}_f
\)
is updated by
\(
\bm{\mathcal S}_f^{+}
=
\bm{\mathcal S}_f
+
\bm\phi_{f,\mathrm{new}}
\bm\phi_{f,\mathrm{new}}^{\top}
\).

\begin{assumption}[Finite excitation]
\label{ass:finite_excitation}
There exist a finite \(M_f\geq N_f\) and a constant
\(\alpha_f>0\) such that the stored integral data satisfy
\(
\bm{\mathcal S}_f
=
\sum_{\ell=0}^{M_f-1}
\bm\phi_{f,\ell}\bm\phi_{f,\ell}^{\top}
\succeq
\alpha_f\bm I
\).
\end{assumption}

In addition to the stored data, define the current integral data over a
moving window of length \(T_f>0\) by
\(
\bm y_f(t)
:=
\bx(t)-\bx(t-T_f)
-
\int_{t-T_f}^{t}
\bg(\bx(\tau))\bu(\tau)\,d\tau
\)
and
\(
\bm\phi_f(t)
:=
\int_{t-T_f}^{t}
\bsigma_f(\bx(\tau))\,d\tau
\).
From \eqref{eq:plant} and \eqref{eq:f_approximation}, these quantities
satisfy
\(
\bm y_f(t)
=
\bW_f^{*\top}\bm\phi_f(t)
+
\bm d_f(t)
\),
where
\(
\bm d_f(t)
:=
\int_{t-T_f}^{t}
\bm{\varepsilon}_f(\bx(\tau))\,d\tau
\).
Similarly, every stored pair satisfies
\(
\bm y_{f,\ell}
=
\bW_f^{*\top}\bm\phi_{f,\ell}
+
\bm d_{f,\ell}
\).

Using the current estimate \(\hat{\bW}_f(t)\), define the current and
stored integral prediction errors as
\(
\bm e_f(t)
:=
\bm y_f(t)
-
\hat{\bW}_f^\top(t)\bm\phi_f(t)
\)
and
\(
\bm e_{f,\ell}(t)
:=
\bm y_{f,\ell}
-
\hat{\bW}_f^\top(t)\bm\phi_{f,\ell}
\),
respectively. Consider the composite prediction-error objective
\(
J_f
:=
\frac{1}{2}\|\bm e_f(t)\|^2
+
\frac{k_f}{2}
\sum_{\ell=0}^{M_f-1}
\|\bm e_{f,\ell}(t)\|^2
\),
where \(k_f>0\). Since
\(
\nabla_{\hat{\bW}_f}J_f
=
-\bm\phi_f(t)\bm e_f^\top(t)
-
k_f
\sum_{\ell=0}^{M_f-1}
\bm\phi_{f,\ell}\bm e_{f,\ell}^\top(t)
\),
the gradient update
\(
\dot{\hat{\bW}}_f
=
-\bm{\Gamma}_f
\nabla_{\hat{\bW}_f}J_f
\)
gives
\begin{equation}
\dot{\hat{\bW}}_f
=
\bm{\Gamma}_f
\left[
    \bm\phi_f(t)\bm e_f^\top(t)
    +
    k_f
    \sum_{\ell=0}^{M_f-1}
    \bm\phi_{f,\ell}\bm e_{f,\ell}^\top(t)
\right].
\label{eq:icl_identifier}
\end{equation}
where
\(
\bm{\Gamma}_f=\bm{\Gamma}_f^\top\succ0
\).
Hence, the identifier combines the current moving-window information
with the informative historical data retained in the history stack.

Let
\(
\tilde{\bW}_f
:=
\hat{\bW}_f-\bW_f^*
\).
The above regression relations immediately yield
\(
\bm e_f
=
-\tilde{\bW}_f^\top\bm\phi_f+\bm d_f
\)
and
\(
\bm e_{f,\ell}
=
-\tilde{\bW}_f^\top\bm\phi_{f,\ell}
+
\bm d_{f,\ell}
\).
Accordingly, the identification dynamics are analyzed with
\(
V_f
:=
\frac{1}{2}
\operatorname{tr}
\left(
\tilde{\bW}_f^\top
\bm{\Gamma}_f^{-1}
\tilde{\bW}_f
\right)
\).

\begin{lemma}[Composite identification convergence]
\label{lem:identification_error}
Suppose that Assumptions~\ref{ass:plant}--\ref{ass:finite_excitation}
hold and that the integral approximation residuals are bounded.
Then the weight-estimation error generated by
\eqref{eq:icl_identifier} is uniformly ultimately bounded. In
particular, there exist constants \(c_f>0\) and \(\Delta_f\geq0\)
such that
\begin{equation}
    \dot V_f
    \leq
    -c_fV_f+\Delta_f .
    \label{eq:icl_weight_bound}
\end{equation}
Consequently, the reconstructed drift
\(
\hat{\bfv}(\bx,t)
=
\hat{\bW}_f^\top(t)\bsigma_f(\bx)
\)
satisfies
\(
\|\bfv(\bx)-\hat{\bfv}(\bx,t)\|
\leq
\bar\sigma_f\|\tilde{\bW}_f(t)\|_F
+
\bar\varepsilon_f
\)
for every
\(
\bx\in\Omega
\).
If
\(
\bm{\varepsilon}_f\equiv\bm0
\),
then
\(
\Delta_f=0
\)
and
\(
\tilde{\bW}_f(t)\to\bm0
\).
\end{lemma}



\begin{proof}
Differentiating \(V_f\) along \eqref{eq:icl_identifier}, and noting that the ideal weight matrix is constant so that
\(\dot{\tilde{\bW}}_f=\dot{\hat{\bW}}_f\), gives
\begin{equation}
\dot V_f
=
\operatorname{tr}
[
\tilde{\bW}_f^\top
\bm\phi_f\bm e_f^\top
]
+
k_f
\sum_{\ell=0}^{M_f-1}
\operatorname{tr}
[
\tilde{\bW}_f^\top
\bm\phi_{f,\ell}
\bm e_{f,\ell}^{\top}
].
\label{eq:Vf_dot_initial}
\end{equation}
From the current regression relation and its stored counterparts, the corresponding prediction errors can be written as
\(
\bm e_f
=
-\tilde{\bW}_f^\top\bm\phi_f+\bm d_f
\)
and
\(
\bm e_{f,\ell}
=
-\tilde{\bW}_f^\top\bm\phi_{f,\ell}
+
\bm d_{f,\ell}
\),
respectively. Substituting these expressions into \eqref{eq:Vf_dot_initial} yields
\begin{equation}
\begin{aligned}
\dot V_f
={}&
-\operatorname{tr}
[
\tilde{\bW}_f^\top
\bm\phi_f
\bm\phi_f^\top
\tilde{\bW}_f
]
-
k_f
\sum_{\ell=0}^{M_f-1}
\operatorname{tr}
[
\tilde{\bW}_f^\top
\bm\phi_{f,\ell}
\bm\phi_{f,\ell}^\top
\tilde{\bW}_f
]
\\
&+
\operatorname{tr}
[
\tilde{\bW}_f^\top
\bm\phi_f
\bm d_f^\top
]
\\
&+
k_f
\sum_{\ell=0}^{M_f-1}
\operatorname{tr}
[
\tilde{\bW}_f^\top
\bm\phi_{f,\ell}
\bm d_{f,\ell}^\top
].
\end{aligned}
\label{eq:Vf_dot_expanded}
\end{equation}
Using
\(
\operatorname{tr}
[
\tilde{\bW}_f^\top
\bm\phi_f
\bm\phi_f^\top
\tilde{\bW}_f
]
=
\|\tilde{\bW}_f^\top\bm\phi_f\|^2
\)
and, similarly,
\(
\operatorname{tr}
[
\tilde{\bW}_f^\top
\bm\phi_{f,\ell}
\bm\phi_{f,\ell}^\top
\tilde{\bW}_f
]
=
\|\tilde{\bW}_f^\top\bm\phi_{f,\ell}\|^2
\),
one obtains
\begin{equation}
\begin{aligned}
\dot V_f
={}&
-\|\tilde{\bW}_f^\top\bm\phi_f\|^2
-
k_f
\sum_{\ell=0}^{M_f-1}
\|\tilde{\bW}_f^\top\bm\phi_{f,\ell}\|^2
\\
&+
\operatorname{tr}
[
\tilde{\bW}_f^\top
\bm\phi_f
\bm d_f^\top
]
+
k_f
\sum_{\ell=0}^{M_f-1}
\operatorname{tr}
[
\tilde{\bW}_f^\top
\bm\phi_{f,\ell}
\bm d_{f,\ell}^\top
].
\end{aligned}
\label{eq:Vf_dot_residual}
\end{equation}

The last two terms in \eqref{eq:Vf_dot_residual} arise from the integral approximation residuals and therefore must be bounded explicitly. By the Cauchy--Schwarz inequality for the Frobenius inner product,
\(
\left|
\operatorname{tr}
[
\tilde{\bW}_f^\top
\bm\phi_f
\bm d_f^\top
]
\right|
\leq
\|\tilde{\bW}_f^\top\bm\phi_f\|
\|\bm d_f\|
\)
and
\(
\left|
\operatorname{tr}
[
\tilde{\bW}_f^\top
\bm\phi_{f,\ell}
\bm d_{f,\ell}^\top
]
\right|
\leq
\|\tilde{\bW}_f^\top\bm\phi_{f,\ell}\|
\|\bm d_{f,\ell}\|
\).
For arbitrary constants
\(
\epsilon_{f,0}\in(0,2)
\)
and
\(
\epsilon_{f,\ell}\in(0,2)
\),
Young's inequality gives
\begin{equation}
\|\tilde{\bW}_f^\top\bm\phi_f\|
\|\bm d_f\|
\leq
\frac{\epsilon_{f,0}}{2}
\|\tilde{\bW}_f^\top\bm\phi_f\|^2
+
\frac{1}{2\epsilon_{f,0}}
\|\bm d_f\|^2,
\label{eq:young_current}
\end{equation}
and
\begin{equation}
\|\tilde{\bW}_f^\top\bm\phi_{f,\ell}\|
\|\bm d_{f,\ell}\|
\leq
\frac{\epsilon_{f,\ell}}{2}
\|\tilde{\bW}_f^\top\bm\phi_{f,\ell}\|^2
+
\frac{1}{2\epsilon_{f,\ell}}
\|\bm d_{f,\ell}\|^2.
\label{eq:young_stored}
\end{equation}
Substituting \eqref{eq:young_current} and \eqref{eq:young_stored} into \eqref{eq:Vf_dot_residual} gives
\begin{equation}
\begin{aligned}
\dot V_f
\leq{}&
-
\left(
1-\frac{\epsilon_{f,0}}{2}
\right)
\|\tilde{\bW}_f^\top\bm\phi_f\|^2
-
k_f
\sum_{\ell=0}^{M_f-1}
\left(
1-\frac{\epsilon_{f,\ell}}{2}
\right)
\|\tilde{\bW}_f^\top\bm\phi_{f,\ell}\|^2
\\
&+
\frac{1}{2\epsilon_{f,0}}
\|\bm d_f\|^2
+
\frac{k_f}{2}
\sum_{\ell=0}^{M_f-1}
\frac{1}{\epsilon_{f,\ell}}
\|\bm d_{f,\ell}\|^2.
\end{aligned}
\label{eq:Vf_dot_young}
\end{equation}

Since the integral approximation residuals are bounded, there exist finite constants
\(
\bar d_f\geq0
\)
and
\(
\bar d_{f,\ell}\geq0
\)
such that
\(
\|\bm d_f\|\leq\bar d_f
\)
and
\(
\|\bm d_{f,\ell}\|\leq\bar d_{f,\ell}
\).
Define
\(
\underline\epsilon_f
:=
\min_{\ell=0,\ldots,M_f-1}
\left(
1-\frac{\epsilon_{f,\ell}}{2}
\right)>0
\).
Then \eqref{eq:Vf_dot_young} implies
\begin{equation}
\begin{aligned}
\dot V_f
\leq{}&
-
\left(
1-\frac{\epsilon_{f,0}}{2}
\right)
\|\tilde{\bW}_f^\top\bm\phi_f\|^2
-
k_f\underline\epsilon_f
\sum_{\ell=0}^{M_f-1}
\|\tilde{\bW}_f^\top\bm\phi_{f,\ell}\|^2
+
\Delta_{f,r},
\end{aligned}
\label{eq:Vf_dot_bounded_residual}
\end{equation}
where
\(
\Delta_{f,r}
:=
\frac{\bar d_f^2}{2\epsilon_{f,0}}
+
\frac{k_f}{2}
\sum_{\ell=0}^{M_f-1}
\frac{\bar d_{f,\ell}^2}{\epsilon_{f,\ell}}
\geq0
\).

By Assumption~\ref{ass:finite_excitation},
\(
\bm{\mathcal S}_f
=
\sum_{\ell=0}^{M_f-1}
\bm\phi_{f,\ell}\bm\phi_{f,\ell}^\top
\succeq
\alpha_f I
\).
Hence,
\begin{equation}
\begin{aligned}
\sum_{\ell=0}^{M_f-1}
\|\tilde{\bW}_f^\top\bm\phi_{f,\ell}\|^2
&=
\sum_{\ell=0}^{M_f-1}
\operatorname{tr}
[
\tilde{\bW}_f^\top
\bm\phi_{f,\ell}
\bm\phi_{f,\ell}^\top
\tilde{\bW}_f
]
=
\operatorname{tr}
[
\tilde{\bW}_f^\top
\bm{\mathcal S}_f
\tilde{\bW}_f
]
\\
&\geq
\alpha_f
\operatorname{tr}
[
\tilde{\bW}_f^\top
\tilde{\bW}_f
]
\\
&=
\alpha_f
\|\tilde{\bW}_f\|_F^2.
\end{aligned}
\label{eq:stored_data_lower_bound}
\end{equation}
Using \eqref{eq:stored_data_lower_bound} in \eqref{eq:Vf_dot_bounded_residual}, and discarding the additional nonpositive current-data term, gives
\begin{equation}
\dot V_f
\leq
-
k_f\underline\epsilon_f\alpha_f
\|\tilde{\bW}_f\|_F^2
+
\Delta_{f,r}.
\label{eq:Vf_dot_weight_norm}
\end{equation}
Thus, defining
\(
\bar c_f
:=
k_f\underline\epsilon_f\alpha_f>0
\)
and
\(
\Delta_f:=\Delta_{f,r}\geq0
\),
one obtains
\begin{equation}
\dot V_f
\leq
-\bar c_f
\|\tilde{\bW}_f\|_F^2
+
\Delta_f.
\label{eq:Vf_dot_intermediate}
\end{equation}

Finally, from the definition
\(
V_f
=
\frac{1}{2}
\operatorname{tr}
[
\tilde{\bW}_f^\top
\bm{\Gamma}_f^{-1}
\tilde{\bW}_f
]
\)
and
\(
\bm{\Gamma}_f\succ0
\),
it follows that
\(
V_f
\leq
[2\lambda_{\min}(\bm{\Gamma}_f)]^{-1}
\|\tilde{\bW}_f\|_F^2
\),
or equivalently
\(
\|\tilde{\bW}_f\|_F^2
\geq
2\lambda_{\min}(\bm{\Gamma}_f)V_f
\).
Substituting this inequality into \eqref{eq:Vf_dot_intermediate} yields
\begin{equation}
\dot V_f
\leq
-
2\bar c_f
\lambda_{\min}(\bm{\Gamma}_f)
V_f
+
\Delta_f.
\label{eq:Vf_dot_final_step}
\end{equation}
Therefore, by defining
\(
c_f
:=
2\bar c_f
\lambda_{\min}(\bm{\Gamma}_f)>0
\),
one obtains
\(
\dot V_f
\leq
-c_fV_f+\Delta_f
\),
which proves \eqref{eq:icl_weight_bound}.
\end{proof}

The online reconstructed model
\(
\hat{\bfv}(\bx,t)
=
\hat{\bW}_f^\top(t)\bsigma_f(\bx)
\)
is subsequently used in the inverse-optimal construction and is
updated concurrently with the closed-loop evolution.

\begin{lemma}[Bounded identification error]
\label{lem:identification_error}
Under Assumptions~\ref{ass:plant}--\ref{ass:finite_excitation},
the identification error is uniformly bounded on \(\Omega\). In particular,
there exists a finite constant \(\bar\Delta_f>0\) such that
\begin{equation}
    \left\|
        \bfv(\bx)-\hat{\bfv}(\bx,t)
    \right\|
    \leq
    \bar\Delta_f,
    \qquad
    \forall\,\bx\in\Omega .
    \label{eq:identification_error_bound}
\end{equation}
\end{lemma}

\begin{proof}
From
\(
\bfv(\bx)
=
\bW_f^{*\top}\bsigma_f(\bx)
+
\bm{\varepsilon}_f(\bx)
\)
and
\(
\hat{\bfv}(\bx,t)
=
\hat{\bW}_f^\top(t)\bsigma_f(\bx)
\),
one has
\(
\bfv(\bx)-\hat{\bfv}(\bx,t)
=
-\tilde{\bW}_f^\top(t)\bsigma_f(\bx)
+
\bm{\varepsilon}_f(\bx)
\).
Hence,
\(
\|\bfv(\bx)-\hat{\bfv}(\bx,t)\|
\leq
\bar\sigma_f\|\tilde{\bW}_f(t)\|_F
+
\bar\varepsilon_f
\).
By Lemma~\ref{lem:identification_error},
\(
\tilde{\bW}_f(t)
\)
is uniformly ultimately bounded. Therefore, there exists a finite
constant \(\bar W_f>0\) such that
\(
\|\tilde{\bW}_f(t)\|_F\leq\bar W_f
\),
and thus \eqref{eq:identification_error_bound} holds with
\(
\bar\Delta_f
:=
\bar\sigma_f\bar W_f
+
\bar\varepsilon_f
\).
\end{proof}


\begin{remark}
Before activating the online update law \eqref{eq:icl_identifier}, the stored integral regressions satisfy \(\bm y_{f,\ell}\approx\hat{\bW}_{f}^{\top}\bm\phi_{f,\ell}\), \(\ell=0,\ldots,M_f-1\). Collecting \(M_f\) data segments gives \(\bm Y_f\approx\hat{\bW}_{f}^{\top}\bm\Phi_f\), where \(\bm Y_f=[\bm y_{f,0},\ldots,\bm y_{f,M_f-1}]\) and \(\bm\Phi_f=[\bm\phi_{f,0},\ldots,\bm\phi_{f,M_f-1}]\). Under Assumption~\ref{ass:finite_excitation}, \(\bm\Phi_f\bm\Phi_f^\top\succ0\), and therefore the initial weight estimate can be obtained from the least-squares solution \(\hat{\bW}_{f,\mathrm{LS}}^\top=\bm Y_f\bm\Phi_f^\top(\bm\Phi_f\bm\Phi_f^\top)^{-1}\), equivalently, \(\hat{\bW}_{f,\mathrm{LS}}^\top=\left(\sum_{\ell=0}^{M_f-1}\bm y_{f,\ell}\bm\phi_{f,\ell}^\top\right)\left(\sum_{\ell=0}^{M_f-1}\bm\phi_{f,\ell}\bm\phi_{f,\ell}^\top\right)^{-1}\). This batch solution is then used as the informed initialization \(\hat{\bW}_f(0)=\hat{\bW}_{f,\mathrm{LS}}\) for the subsequent online update. Nevertheless, because the integral regression is only approximate, the batch estimate should not be interpreted as the exact ideal weight \(\bW_f^*\); rather, it provides a data-informed starting point for the online identifier. This distinction is important because an arbitrary trial-and-error initialization may produce a large initial parameter mismatch \(\tilde{\bW}_f(0)=\hat{\bW}_f(0)-\bW_f^*\), whereas the least-squares initialization exploits the informative history stack before online adaptation begins. The subsequent update \eqref{eq:icl_identifier} then combines the current moving-window regression with the stored informative data to refine the weight estimate online, following the concurrent/integral-concurrent-learning philosophy in \citep{Chowdhary2010CL,Parikh2019ICL}.
\end{remark}


\section{Fixed-Time Reinforcement Learning Optimal Control}
\label{sec:fixed_time_rl}


\subsection{Fixed-Time Optimal-Control Formulation}

The fixed-time comparison result in Theorem~\ref{thm:fixed_time_stability} provides a direct mechanism for guaranteeing a uniform upper bound on the settling time independently of the initial conditions. In particular, rather than assigning a convergence deadline in advance, a designer may select positive decay gains and suitable exponents to obtain an explicit upper bound on the settling time. Motivated by this observation, we first select a continuously differentiable positive-definite function \(V_d:\Omega\rightarrow\mathbb R_{\geq0}\) with \(V_d(\bm 0)=0\), and prescribe its desired closed-loop evolution as
\begin{equation}
\dot V_d(\bx)
=
-k_1V_d^{\gamma_1}(\bx)
-k_2V_d^{\frac{\gamma_2+1}{2}}(\bx),
\label{eq:desired_fixed_time_decay}
\end{equation}
where \(k_1,k_2>0\) and \(\frac12<\gamma_1<1<\gamma_2\).

According to Theorem~\ref{thm:fixed_time_stability}, the corresponding settling-time upper bound is given by
\begin{equation}
T(\bx_0)
\leq
T_{\max}
:=
\frac{1}{k_1(1-\gamma_1)}
+
\frac{2}{k_2(\gamma_2-1)}.
\label{eq:fixed_time_state_bound}
\end{equation}

The function \(V_d\) is freely selected by the designer to encode the desired closed-loop behavior. According to Theorem~\ref{thm:fixed_time_stability}, any feedback law satisfying \eqref{eq:desired_fixed_time_decay} renders the origin fixed-time stable and guarantees \(T(\bx_0)\leq T_{\max}\), independently of the initial conditions, provided that the corresponding trajectories remain in the domain where the decay condition holds. Hence, \(T_{\max}\) is determined by the selected decay gains and exponents rather than being prescribed independently before the controller is constructed. The remaining development therefore seeks an optimal feedback law whose closed-loop derivative of \(V_d\) exactly reproduces \eqref{eq:desired_fixed_time_decay}.

\begin{remark}

Here, \(T_{\max}\) is understood as the fixed-time
convergence bound of the system once the optimal
value function has been obtained. However, the overall
convergence time also involves data collection for
system identification and the time required for the
critic to learn the optimal value function.
Although these processes may not occur sequentially
in practice, the states may approach the origin even
while the critic weights are still being learned.
Nevertheless, such a decomposition represents a
theoretical perspective for establishing convergence
guarantees, which remain valid regardless of the
actual temporal overlap among these processes.
Assume that the duration of data collection for
identification is prescribed in advance and is
independent of the initial conditions.
Excluding this identification data-collection period,
the remainder of this paper considers only the
critic-learning time and the closed-loop convergence
time.

\end{remark}


Consider the performance index
~\cite{Kokolakis2025Predefined}
\begin{equation}
\begin{aligned}
J(\bx_0,\bu)
={}&
\int_0^\infty
\left[
    \bm L_1(\bx)\bu
    +
    L_2(\bx)
    +
    \bu^\top\bR\bu
\right]dt,
\qquad
\bR=\bR^\top>0.
\end{aligned}
\label{eq:cost_function}
\end{equation}

Here, \(\bm L_1:\Omega\to\mathbb R^{1\times m}\)
and \(L_2:\Omega\to\mathbb R\) are to be
constructed such that \(V_d\) is the optimal
value function and the corresponding policy
satisfies the prescribed fixed-time decay
\eqref{eq:desired_fixed_time_decay}.
The term \(\bu^\top\bR\bu\) penalizes control
effort, while \(\bm L_1(\bx)\bu\) introduces
state-control coupling and \(L_2(\bx)\)
penalizes state deviations when nonnegative
\cite{Kokolakis2025Predefined}.
The running cost need not be positive definite
for every state-input pair.
As discussed in
\cite[Sec.~8.4, pp.~524--525]{Haddad2008NonlinearControl},
the state-control cross-weighting term may
render the running cost negative for certain
control inputs without necessarily
invalidating inverse optimality over the
class of admissible stabilizing policies.
In particular, inverse optimality can be
established through the Hamilton--Jacobi--Bellman
equation and the corresponding admissibility
and transversality conditions, even when
the running cost is not pointwise nonnegative.
A detailed verification of this property
for the present inverse-optimal construction
is provided in~\ref{app:running_cost_verification}.
The constructed coefficients satisfy
\(\bm L_1(\bm0)=\bm0\) and \(L_2(\bm0)=0\),
as established subsequently (See Lemma~\ref{lem:continuous_extension}).

For the identified model \(\dot{\bx}=\hat{\bfv}(\bx)+\bg(\bx)\bu\), the Hamiltonian associated with \eqref{eq:cost_function} is
\begin{equation}
\begin{aligned}
H(\bx,\bu,\nabla V_d)
={}&
\bm L_1(\bx)\bu
+
L_2(\bx)
+
\bu^\top\bR\bu
\\
&+
\nabla V_d^\top(\bx)
\hat{\bfv}(\bx)
+
\nabla V_d^\top(\bx)
\bg(\bx)\bu .
\end{aligned}
\label{eq:hamiltonian_predefined_time}
\end{equation}

Equivalently, \(H=L_2+\nabla V_d^\top\hat{\bfv}+\bu^\top\bR\bu+\left(\bm L_1+\nabla V_d^\top\bg\right)\bu\). The corresponding HJB equation is \(\min_{\bu}H(\bx,\bu,\nabla V_d)=0\). Since \(\bR>0\), one has \(\partial^2H/\partial\bu^2=2\bR>0\), and therefore the unique minimizing policy satisfies \(\partial H/\partial\bu=\bm L_1^\top(\bx)+2\bR\bu+\bg^\top(\bx)\nabla V_d(\bx)=\bm0\). Hence, \(2\bR\bu^*=-\left[\bm L_1^\top(\bx)+\bg^\top(\bx)\nabla V_d(\bx)\right]\), which gives
\begin{equation}
\bu^*(\bx)
=
-\frac{1}{2}
\bR^{-1}
\left[
    \bm L_1^\top(\bx)
    +
    \bg^\top(\bx)\nabla V_d(\bx)
\right].
\label{eq:optimal_control_general}
\end{equation}


\subsection{Reverse Construction of the Fixed-Time Optimal Cost}
\label{subsec:reverse_fixed_time_cost}


The prescribed predefined-time decay is embedded into an inverse-optimal
minimum-control-effort design. This construction is consistent with the
inverse-optimal control framework developed by Haddad and Chellaboina
~\cite[Sec.~8.4, pp.~522--524]{Haddad2008NonlinearControl}, in which
a stabilizing feedback law is prescribed and a corresponding performance
functional is constructed to establish its optimality.
Additional feasibility and small-control
conditions are introduced ~\cite{Sontag1989Universal} to ensure that the resulting feedback is well
defined and admits a continuous extension at the origin. Such feasibility
and continuous-extension issues are not explicitly addressed in the
predefined-time inverse-optimal formulation of
Kokolakis \emph{et al.}~\cite{Kokolakis2025Predefined}, which motivates
their explicit treatment here.


\begin{theorem}[Inverse-optimal fixed-time stabilization]
\label{thm:inverse_optimal_pt}
Suppose that Assumptions \ref{ass:plant}--\ref{ass:inverse_optimal_feasibility} hold. For the minimum-control-effort
construction, choose
\(\bm L_1(\bx)
=
\left[
\frac{
2\left(
\nabla V_d^\top(\bx)\hat{\bfv}(\bx)
+
k_1V_d^{\gamma_1}(\bx)+k_2V_d^{\frac{\gamma_2+1}{2}}(\bx)
\right)
}{
\nabla V_d^\top(\bx)
\bg(\bx)
\bR^{-1}
\bg^\top(\bx)
\nabla V_d(\bx)
}
-1
\right]
\nabla V_d^\top(\bx)\bg(\bx)\)
and
\(L_2(\bx)
=
-\nabla V_d^\top(\bx)\hat{\bfv}(\bx)
+
\frac{
\left(
\nabla V_d^\top(\bx)\hat{\bfv}(\bx)
+
k_1V_d^{\gamma_1}(\bx)+k_2V_d^{\frac{\gamma_2+1}{2}}(\bx)
\right)^2
}{
\nabla V_d^\top(\bx)
\bg(\bx)
\bR^{-1}
\bg^\top(\bx)
\nabla V_d(\bx)
}\).



Then, the running cost
\(L(\bx,\bu)
=
\bm L_1(\bx)\bu
+
L_2(\bx)
+
\bu^\top\bR\bu\)
admits the inverse-optimal feedback policy
\(\bu_c(\bx)\).
Moreover, the closed-loop identified system
\(\dot{\bx}
=
\hat{\bfv}(\bx)
+
\bg(\bx)\bu_c(\bx)\)
satisfies
\(\dot V_d(\bx)
\leq
-k_1V_d^{\gamma_1}(\bx)
-k_2V_d^{\frac{\gamma_2+1}{2}}(\bx)\),
and hence, by
Theorem~\ref{thm:fixed_time_stability},
is fixed-time stable on any forward-invariant
domain where the stated conditions hold, with
\(T(\bx_0)\leq T_{d,\max}\), where
\(T_{d,\max}
=
1/[k_1(1-\gamma_1)]
+
2/[k_2(\gamma_2-1)]\).
\end{theorem}

\begin{proof}

Along the identified dynamics, the derivative of \(V_d\) is \(\dot V_d=\nabla V_d^\top(\bx)\hat{\bfv}(\bx)+\nabla V_d^\top(\bx)\bg(\bx)\bu\). Imposing the prescribed fixed-time evolution \eqref{eq:desired_fixed_time_decay} yields
\[
\nabla V_d^\top(\bx)\hat{\bfv}(\bx)
+
\nabla V_d^\top(\bx)\bg(\bx)\bu
=
-k_1V_d^{\gamma_1}(\bx)-k_2V_d^{\frac{\gamma_2+1}{2}}(\bx).
\]
Equivalently, \(\nabla V_d^\top(\bx)\bg(\bx)\bu=-\sigma_d(\bx)\), where
\begin{equation}
\begin{aligned}
\sigma_d(\bx)
:={}&
\nabla V_d^\top(\bx)\hat{\bfv}(\bx)
+
k_1V_d^{\gamma_1}(\bx)+k_2V_d^{\frac{\gamma_2+1}{2}}(\bx).
\end{aligned}
\label{eq:sigma_d_fixed}
\end{equation}
Define also
\begin{equation}
D_d(\bx)
:=
\nabla V_d^\top(\bx)
\bg(\bx)
\bR^{-1}
\bg^\top(\bx)
\nabla V_d(\bx).
\label{eq:Dd_fixed}
\end{equation}

Since infinitely many control inputs may satisfy \(\nabla V_d^\top(\bx)\bg(\bx)\bu+\sigma_d(\bx)=0\), we select the one minimizing the weighted control effort \(\bu^\top\bR\bu\). Accordingly, consider \(\min_{\bu}\bu^\top\bR\bu\) subject to \(\nabla V_d^\top(\bx)\bg(\bx)\bu+\sigma_d(\bx)=0\). The corresponding Lagrangian is \(\mathcal L_c(\bu,\lambda_c)=\bu^\top\bR\bu+\lambda_c\left[\nabla V_d^\top(\bx)\bg(\bx)\bu+\sigma_d(\bx)\right]\). The stationarity condition with respect to \(\bu\) gives \(\partial\mathcal L_c/\partial\bu=2\bR\bu+\lambda_c\bg^\top(\bx)\nabla V_d(\bx)=\bm0\), and therefore \(\bu=-(\lambda_c/2)\bR^{-1}\bg^\top(\bx)\nabla V_d(\bx)\). Substituting this expression into the constraint yields \(-(\lambda_c/2)D_d(\bx)+\sigma_d(\bx)=0\), so that \(\lambda_c=2\sigma_d(\bx)/D_d(\bx)\). Hence, the unique \(\bR\)-weighted minimum-effort feedback satisfying the prescribed fixed-time decay is

\begin{equation}
\begin{aligned}
\bu_c(\bx)
={}&
-
\frac{
\nabla V_d^\top(\bx)\hat{\bfv}(\bx)
+
k_1V_d^{\gamma_1}(\bx)+k_2V_d^{\frac{\gamma_2+1}{2}}(\bx)
}{
\nabla V_d^\top(\bx)
\bg(\bx)\bR^{-1}
\bg^\top(\bx)
\nabla V_d(\bx)
}
\bR^{-1}
\bg^\top(\bx)
\nabla V_d(\bx).
\end{aligned}
\label{eq:continuous_fixed_time_controller}
\end{equation}

The following condition ensures that \eqref{eq:continuous_fixed_time_controller} is well defined.

\begin{assumption}[Inverse-optimal feasibility]
\label{ass:inverse_optimal_feasibility}
For every \(\bx\in\Omega\setminus\{\bm0\}\), \(\nabla V_d^\top(\bx)\bg(\bx)\bR^{-1}\bg^\top(\bx)\nabla V_d(\bx)>0\).
\end{assumption}

\begin{remark}
Assumption~\ref{ass:inverse_optimal_feasibility} excludes nonzero states
at which \(D_d(\bm{x})=0\), equivalently
\(\bm{g}^\top(\bm{x})\nabla V_d(\bm{x})=\bm{0}\).
For example, if
\(\bm{g}=[1\;0]^\top\) and
\(V_d(\bm{x})=\frac{1}{2}(x_1^2+x_2^2)\),
then \(D_d(\bm{x})=0\) whenever \(x_1=0\), even if \(x_2\neq0\).
\end{remark}

Indeed, using \eqref{eq:continuous_fixed_time_controller}, one obtains \(\nabla V_d^\top\bg\bu_c=-(\sigma_d/D_d)\nabla V_d^\top\bg\bR^{-1}\bg^\top\nabla V_d=-\sigma_d\), and therefore
\[
\dot V_d
=
\nabla V_d^\top\hat{\bfv}
+
\nabla V_d^\top\bg\bu_c
=
\nabla V_d^\top\hat{\bfv}
-
\sigma_d
=
-k_1V_d^{\gamma_1}-k_2V_d^{\frac{\gamma_2+1}{2}}.
\]

We now construct \(\bm L_1(\bx)\) such that the Hamiltonian-minimizing policy \(\bu^*\) coincides with \(\bu_c\). From \eqref{eq:optimal_control_general}, \(\bu^*=-(1/2)\bR^{-1}\left[\bm L_1^\top+\bg^\top\nabla V_d\right]\), whereas \(\bu_c=-(\sigma_d/D_d)\bR^{-1}\bg^\top\nabla V_d\). Imposing \(\bu^*=\bu_c\) gives \(\bm L_1^\top+\bg^\top\nabla V_d=(2\sigma_d/D_d)\bg^\top\nabla V_d\), so that
\begin{equation}
\bm L_1^\top(\bx)
=
\left[
    \frac{2\sigma_d(\bx)}{D_d(\bx)}
    -
    1
\right]
\bg^\top(\bx)\nabla V_d(\bx).
\label{eq:L1_design_compact_fixed}
\end{equation}
Substituting \eqref{eq:sigma_d_fixed} and \eqref{eq:Dd_fixed} gives the explicit expression

\begin{equation}
\begin{aligned}
\bm L_1^\top(\bx)
={}&
\left[
\frac{
    2
    \left[
        \nabla V_d^\top(\bx)\hat{\bfv}(\bx)
        +
        k_1V_d^{\gamma_1}(\bx)+k_2V_d^{\frac{\gamma_2+1}{2}}(\bx)
    \right]
}{
    \nabla V_d^\top(\bx)
    \bg(\bx)
    \bR^{-1}
    \bg^\top(\bx)
    \nabla V_d(\bx)
}
-
1
\right]
\\
&\qquad\times
\bg^\top(\bx)\nabla V_d(\bx).
\end{aligned}
\label{eq:L1_design}
\end{equation}

It remains to determine \(L_2(\bx)\). Substitution of \eqref{eq:optimal_control_general} into the Hamiltonian gives \(H^*=L_2+\nabla V_d^\top\hat{\bfv}-\frac{1}{4}\left[\bm L_1^\top+\bg^\top\nabla V_d\right]^\top\bR^{-1}\left[\bm L_1^\top+\bg^\top\nabla V_d\right]\). Using \(\bm L_1^\top+\bg^\top\nabla V_d=(2\sigma_d/D_d)\bg^\top\nabla V_d\), the quadratic term becomes
\[
\frac{1}{4}
\left[
    \bm L_1^\top+\bg^\top\nabla V_d
\right]^\top
\bR^{-1}
\left[
    \bm L_1^\top+\bg^\top\nabla V_d
\right]
=
\frac{\sigma_d^2(\bx)}{D_d(\bx)}.
\]
Hence, \(H^*=L_2+\nabla V_d^\top\hat{\bfv}-\sigma_d^2/D_d\). The HJB condition \(H^*=0\) therefore gives
\begin{equation}
L_2(\bx)
=
-
\nabla V_d^\top(\bx)\hat{\bfv}(\bx)
+
\frac{\sigma_d^2(\bx)}{D_d(\bx)}.
\label{eq:L2_design_compact_fixed}
\end{equation}
Equivalently,

\begin{equation}
\begin{aligned}
L_2(\bx)
={}&
-\nabla V_d^\top(\bx)\hat{\bfv}(\bx)
+
\frac{
\left[
    \nabla V_d^\top(\bx)\hat{\bfv}(\bx)
    +
    k_1V_d^{\gamma_1}(\bx)+k_2V_d^{\frac{\gamma_2+1}{2}}(\bx)
\right]^2
}{
    \nabla V_d^\top(\bx)
    \bg(\bx)
    \bR^{-1}
    \bg^\top(\bx)
    \nabla V_d(\bx)
}.
\end{aligned}
\label{eq:L2_design}
\end{equation}

Hence the Proof complete.
\end{proof}

\begin{lemma}[Sufficient small-control condition]
\label{lem:small_control_condition}
Suppose that, in a neighborhood of the origin,
\(c_1\|\bm{x}\|^m
\leq V_d(\bm{x})
\leq c_2\|\bm{x}\|^m\)
and
\(\|\nabla V_d(\bm{x})\|
\leq c_3\|\bm{x}\|^{m-1}\),
for some \(m>1\).
Under Assumption~\ref{ass:basis_origin}, together with the boundedness
of the identifier weights, there exists a constant \(c_f>0\) such that
\(\|\hat{\bm f}(\bm{x})\|
\leq c_f\|\bm{x}\|\)
in a neighborhood of the origin.
Assume further that there exists a constant \(c_b>0\) such that
\(\|\bm g^\top(\bm{x})\nabla V_d(\bm{x})\|
\geq c_b\|\bm{x}\|^{m-1}\)
for all sufficiently small \(\bm{x}\neq\bm{0}\).
Since
\(D_d(\bm{x})
=
\bigl[\bm g^\top(\bm{x})\nabla V_d(\bm{x})\bigr]^\top
\bm R^{-1}
\bigl[\bm g^\top(\bm{x})\nabla V_d(\bm{x})\bigr]\),
there exists \(c_g>0\), specifically
\(c_g=\lambda_{\min}(\bm R^{-1})c_b^2\),
such that
\(D_d(\bm{x})
\geq c_g\|\bm{x}\|^{2m-2}\)
for all sufficiently small \(\bm{x}\neq\bm{0}\).
If \(m\gamma_1>m-1\), then
\(\lim_{\bm{x}\to\bm{0}}
\frac{|\sigma_d(\bm{x})|}
{\sqrt{D_d(\bm{x})}}=0\).
\end{lemma}

\begin{proof}
From Assumption~\ref{ass:basis_origin} and the boundedness of the
identifier weights, there exists a constant \(c_f>0\) such that
\(\|\hat{\bm f}(\bm{x})\|
\leq c_f\|\bm{x}\|\)
in a neighborhood of the origin.
Hence, using
\(\|\nabla V_d(\bm{x})\|
\leq c_3\|\bm{x}\|^{m-1}\),
one obtains
\(\left|
\nabla V_d^\top(\bm{x})\hat{\bm f}(\bm{x})
\right|
\leq c_3c_f\|\bm{x}\|^m
=\mathcal O(\|\bm{x}\|^m)\).

Moreover, since
\(\gamma_1<\frac{\gamma_2+1}{2}\)
and
\(V_d(\bm{x})=\mathcal O(\|\bm{x}\|^m)\)
near the origin,
\(k_1V_d^{\gamma_1}(\bm{x})
+k_2V_d^{\frac{\gamma_2+1}{2}}(\bm{x})
=\mathcal O\!\left(
\|\bm{x}\|^{m\gamma_1}
\right)\)
as \(\bm{x}\to\bm{0}\).
Therefore,
\(|\sigma_d(\bm{x})|
=\mathcal O(\|\bm{x}\|^m)
+\mathcal O\!\left(
\|\bm{x}\|^{m\gamma_1}
\right)\).

Since
\(\sqrt{D_d(\bm{x})}
\geq \sqrt{c_g}\|\bm{x}\|^{m-1}\),
it follows that
\(\frac{|\sigma_d(\bm{x})|}
{\sqrt{D_d(\bm{x})}}
=\mathcal O(\|\bm{x}\|)
+\mathcal O\!\left(
\|\bm{x}\|^{m\gamma_1-(m-1)}
\right)\).

Because \(m>1\) and \(m\gamma_1>m-1\),
both terms vanish as \(\bm{x}\to\bm{0}\).
Hence,
\(\lim_{\bm{x}\to\bm{0}}
\frac{|\sigma_d(\bm{x})|}
{\sqrt{D_d(\bm{x})}}=0\).
\end{proof}

\begin{lemma}[Continuous extension at the origin]
\label{lem:continuous_extension}
Under Assumption~\ref{ass:inverse_optimal_feasibility}
and the preceding regularity conditions, let the conditions of
Lemma~\ref{lem:small_control_condition} hold. Then the control law
\(\bm{u}_c\) and the inverse-optimal cost coefficients
\(\bm{L}_1\) and \(L_2\) admit continuous extensions at the origin given by
\[
\bm{u}_c(\bm{0})=\bm{0},\qquad
\bm{L}_1(\bm{0})=\bm{0},\qquad
L_2(\bm{0})=0.
\]
\end{lemma}

\begin{proof}
Assumption~\ref{ass:inverse_optimal_feasibility} ensures that
\(D_d(\bm{x})>0\) for
\(\bm{x}\in\Omega\setminus\{\bm{0}\}\), so the expressions
for \(\bm{u}_c(\bm{x})\), \(\bm{L}_1(\bm{x})\), and
\(L_2(\bm{x})\) are well defined away from the origin.
Their continuity there follows from the preceding regularity
conditions. It therefore remains to establish their limits at the origin.

By Lemma~\ref{lem:small_control_condition},
\(\lim_{\bm{x}\to\bm{0}}
\frac{|\sigma_d(\bm{x})|}{\sqrt{D_d(\bm{x})}}
=0\).

First, substituting
\(\bm{u}_c(\bm{x})
=-(\sigma_d(\bm{x})/D_d(\bm{x}))
\bm{R}^{-1}\bm{g}^\top(\bm{x})
\nabla V_d(\bm{x})\)
into its weighted quadratic norm gives
\(\bm{u}_c^\top(\bm{x})\bm{R}\bm{u}_c(\bm{x})
=
\frac{\sigma_d^2(\bm{x})}{D_d^2(\bm{x})}
\nabla V_d^\top(\bm{x})\bm{g}(\bm{x})
\bm{R}^{-1}\bm{g}^\top(\bm{x})
\nabla V_d(\bm{x})
=
\frac{\sigma_d^2(\bm{x})}{D_d(\bm{x})}\).
Since \(\bm{R}\succ0\),
\(\lambda_{\min}(\bm{R})\|\bm{u}_c(\bm{x})\|^2
\leq
\bm{u}_c^\top(\bm{x})\bm{R}\bm{u}_c(\bm{x})\),
and consequently
\(\|\bm{u}_c(\bm{x})\|
\leq
\frac{|\sigma_d(\bm{x})|}
{\sqrt{\lambda_{\min}(\bm{R})D_d(\bm{x})}}\).
Hence, Lemma~\ref{lem:small_control_condition} implies
\(\lim_{\bm{x}\to\bm{0}}\bm{u}_c(\bm{x})=\bm{0}\),
which proves continuity of the extension
\(\bm{u}_c(\bm{0})=\bm{0}\).

Next, multiplying the control law by \(\bm{R}\) yields
\(\bm{R}\bm{u}_c(\bm{x})
=
-\frac{\sigma_d(\bm{x})}{D_d(\bm{x})}
\bm{g}^\top(\bm{x})\nabla V_d(\bm{x})\).
Hence, the expression for \(\bm{L}_1\) can be rewritten as
\(\bm{L}_1^\top(\bm{x})
=
\left[
\frac{2\sigma_d(\bm{x})}{D_d(\bm{x})}-1
\right]
\bm{g}^\top(\bm{x})\nabla V_d(\bm{x})
=
-2\bm{R}\bm{u}_c(\bm{x})
-\bm{g}^\top(\bm{x})\nabla V_d(\bm{x})\).
Because \(V_d\) is continuously differentiable and has a minimum
at the origin, \(\nabla V_d(\bm{0})=\bm{0}\) and
\(\nabla V_d(\bm{x})\to\bm{0}\) as \(\bm{x}\to\bm{0}\).
Continuity of \(\bm{g}\) implies that it is locally bounded. Therefore,
\(\|\bm{L}_1(\bm{x})\|
\leq
2\|\bm{R}\|\|\bm{u}_c(\bm{x})\|
+
\|\bm{g}(\bm{x})\|\|\nabla V_d(\bm{x})\|
\to0\).
Thus, \(\bm{L}_1(\bm{0})=\bm{0}\) defines a continuous extension.

Finally, using the quadratic identity above,
\(L_2(\bm{x})
=
-\nabla V_d^\top(\bm{x})\hat{\bm{f}}(\bm{x})
+
\bm{u}_c^\top(\bm{x})\bm{R}\bm{u}_c(\bm{x})\).
By Assumption~\ref{ass:basis_origin} and the boundedness of the
identifier weights, \(\hat{\bm{f}}(\bm{x})\) is locally bounded
near the origin. Hence,
\(|L_2(\bm{x})|
\leq
\|\nabla V_d(\bm{x})\|
\|\hat{\bm{f}}(\bm{x})\|
+
\|\bm{R}\|\|\bm{u}_c(\bm{x})\|^2
\to0\).
Therefore, \(L_2(\bm{0})=0\) also defines a continuous extension,
which completes the proof.
\end{proof}

\begin{corollary}[Design rule for the desired Lyapunov function]
\label{cor:vd_design}
Under the conditions of Lemma~\ref{lem:small_control_condition},
a sufficient design condition ensuring the small-control property is
\(m\gamma_1>m-1\), where \(m>1\) denotes the local growth order of
\(V_d\) around the origin. Combining this condition with the
fixed-time requirement \(0<\gamma_1<1<(\gamma_2+1)/2\) yields the admissible design
range \((m-1)/m<\gamma_1<1<\gamma_2\). In particular, if
\(V_d(\bm{x})=\frac{1}{2}\bm{x}^\top\bm{P}\bm{x}\), with
\(\bm{P}=\bm{P}^\top\succ0\), then \(m=2\), and the design condition
reduces to \(\frac{1}{2}<\gamma_1<1<\gamma_2\).
\end{corollary}

\subsection{Critic-Only Integral Reinforcement Learning}
\label{subsec:critic_only_irl}

\begin{remark}
Although the unknown drift \(\bfv(\bx)\) has been reconstructed in the
preceding identification stage, the estimate \(\hat{\bfv}(\bx)\) is not
treated as an exact model of the plant. It is used only as an auxiliary
quantity for constructing the inverse-optimal terms \(\bm L_1(\bx)\) and
\(L_2(\bx)\) in the preceding section. In the present IRL formulation,
\(\bfv(\bx)\) is still regarded as unknown, and the optimal value function
and control policy are learned directly from integral data without
requiring the exact drift dynamics. Another point to note is that, since the function $V_d$ and its desired behavior are defined with respect to \(\hat{\bfv}(\bx)\) rather than  \(\bfv(\bx)\), the resulting dynamics on  \(\bfv(\bx)\) will only achieve practical fixed-time convergence (see Theorem~\ref{thm:closed_loop}). The discrepancy between \(\bfv(\bx)\) and \(\hat{\bfv}(\bx)\) represented by the term \(\bar\Delta_f\) contributes to the radius of the convergence region (as shown in the proof of Theorem~\ref{thm:closed_loop}).
\end{remark}

After constructing \(\bm L_1(\bx)\) and \(L_2(\bx)\), define
\(\ell(\bx,\bu):=\bm L_1(\bx)\bu+L_2(\bx)+\bu^\top\bR\bu\).
The optimal value function is
\begin{equation}
    V^*(\bx_0)
    =
    \min_{\bu}
    \int_0^\infty
    \ell(\bx(t),\bu(t))\,dt .
    \label{eq:optimal_value_rl}
\end{equation}
For \(\dot{\bx}=\bfv(\bx)+\bg(\bx)\bu\), the Hamiltonian is
\(H(\bx,\bu,\nabla V^*)=\ell(\bx,\bu)+
\nabla V^{*\top}(\bx)[\bfv(\bx)+\bg(\bx)\bu]\).
The HJB equation satisfies
\(0=\min_{\bu}H(\bx,\bu,\nabla V^*)\).
Since \(\partial^2H/\partial\bu^2=2\bR>0\), the minimizing policy follows
from
\(\partial H/\partial\bu=
\bm L_1^\top(\bx)+2\bR\bu+
\bg^\top(\bx)\nabla V^*(\bx)=\bm0\).
Hence,
\begin{equation}
    \bu^*(\bx)
    =
    -\frac{1}{2}
    \bR^{-1}
    \left[
        \bm L_1^\top(\bx)
        +
        \bg^\top(\bx)\nabla V^*(\bx)
    \right].
    \label{eq:critic_only_optimal_policy}
\end{equation}

On the compact set \(\Omega\), let
\begin{equation}
    V^*(\bx)
    =
    \bW^{*\top}\bphi(\bx)
    +
    \varepsilon(\bx),
    \label{eq:critic_value_approximation}
\end{equation}
where \(\bphi(\bx)\in\mathbb R^{N_c}\),
\(\bW^*\in\mathbb R^{N_c}\), and
\(\varepsilon(\bx)\) denotes the approximation error. Then
\(\nabla V^*(\bx)=
(\nabla\bphi(\bx))^\top\bW^*
+
\nabla\varepsilon(\bx)\).
The critic estimate is
\begin{equation}
    \hat V(\bx)
    =
    \hat{\bW}^{\top}\bphi(\bx),
    \qquad
    \nabla\hat V(\bx)
    =
    (\nabla\bphi(\bx))^\top\hat{\bW}.
    \label{eq:critic_estimate}
\end{equation}
Thus, the critic-generated policy is
\begin{equation}
    \hat{\bu}(\bx,\hat{\bW})
    =
    -\frac{1}{2}
    \bR^{-1}
    \left[
        \bm L_1^\top(\bx)
        +
        \bg^\top(\bx)
        (\nabla\bphi(\bx))^\top
        \hat{\bW}
    \right].
    \label{eq:critic_only_policy}
\end{equation}

\begin{proposition}[Critic approximation residual~\cite{Vamvoudakis2010,Kokolakis2025Predefined}]
\label{prop:critic_approximation_residual}
On the compact set
\(
\Omega
\),
suppose that the critic approximation converges
uniformly in the
\(
C^1
\)
sense as
\(
N_C\to\infty
\).
Then,
\[
\lim_{N_C\to\infty}
\sup_{\bx\in\Omega}
\left(
|\varepsilon(\bx)|
+
\|\nabla\varepsilon(\bx)\|
\right)
=0.
\]
\end{proposition}

\begin{remark}[Approximation accuracy and computational trade-off]
Proposition~\ref{prop:critic_approximation_residual}
provides one possible means of reducing the
ultimate bounds established in
Theorems~\ref{thm:fixed_time_critic}
and~\ref{thm:closed_loop}
by increasing the number of critic neurons
\(
N_C
\).
However, approximation residuals are only one
of several factors determining the UUB radii.
Increasing
\(
N_C
\)
may improve approximation accuracy at the
expense of computational complexity, whereas
a smaller network may sacrifice approximation
capability~\cite{ParkSandberg1991}.
This trade-off suggests a possible extension
using an adaptive network architecture, in which
\(
N_C
\)
is adjusted online through neuron allocation
or growing-and-pruning mechanisms, as explored
in~\cite{Platt1991RAN,Han2010SORBF}.
\end{remark}

We have $\hat{\bu}(\bx,\hat{\bW})$ is  the current
learned policy and let $\bu_b(t)$ the behavior input
used to generate the measured trajectory.
During learning, the behavior input satisfies
\begin{equation}
\bu_b(t)
=
\hat{\bu}(\bx(t),\hat{\bW}(t))
+
\be_p(t),
\label{eq:learned_behavior_policy}
\end{equation}
where $\be_p(t)\in\mathbb R^m$ is a bounded
probing signal introduced to provide sufficient
excitation during learning.

The deviation of the behavior input from the
optimal policy can be decomposed as
$\bu_b-\bu^*
=
(\bu_b-\hat{\bu})
+
(\hat{\bu}-\bu^*)
=
\be_p+
(\hat{\bu}-\bu^*)$.

Along the measured dynamics
$\dot{\bx}=\bfv(\bx)+\bg(\bx)\bu_b$,
the optimal Bellman equation is
$0
=
\ell(\bx,\bu^*)
+
\nabla V^{*\top}(\bx)
\left[
\bfv(\bx)+\bg(\bx)\bu^*
\right]$.

Since the measured trajectory is generated
under $\bu_b$ rather than $\bu^*$,
define the policy-suboptimality residual
associated with the current learned policy as
$\delta_{\hat u}(\bx)
:=
\ell(\bx,\hat{\bu})
-
\ell(\bx,\bu^*)
+
\nabla V^{*\top}(\bx)\bg(\bx)
\left[
\hat{\bu}-\bu^*
\right]$.

Consequently,
$\nabla V^{*\top}
\left[
\bfv+\bg\hat{\bu}
\right]
=
-\ell(\bx,\hat{\bu})
+
\delta_{\hat u}(\bx)$.

Using
$\bfv+\bg\bu_b
=
\bfv+\bg\hat{\bu}
+
\bg(\bu_b-\hat{\bu})$,
one obtains
$\dot V^*
=
\nabla V^{*\top}
\left[
\bfv+\bg\bu_b
\right]
=
-\ell(\bx,\hat{\bu})
+
\delta_{\hat u}(\bx)
+
\nabla V^{*\top}\bg
\left(
\bu_b-\hat{\bu}
\right)$.

Thus, the probing signal enters the Bellman
relation through
$\nabla V^{*\top}\bg\be_p$,
whereas $\delta_{\hat u}$ accounts for the
suboptimality of the learned policy.
After the data-excitation phase
\cite{Guo2020CompositeLearning,Parikh2019IntegralCL},
the probing signal is removed, yielding
$\be_p=\mathbf0$.
In contrast, $\delta_{\hat u}=0$ when
$\hat{\bu}=\bu^*$.

Integrating over $[t-T,t]$ gives
\begin{equation}
\begin{aligned}
0
={}&
V^*(\bx(t))
-
V^*(\bx(t-T))
+
\int_{t-T}^{t}
\ell\!\left(
    \bx(\tau),
    \hat{\bu}(\tau)
\right)d\tau
\\
&-
\int_{t-T}^{t}
\nabla V^{*\top}(\bx(\tau))
\bg(\bx(\tau))
\left[
    \bu_b(\tau)
    -
    \hat{\bu}(\tau)
\right]d\tau
-
\int_{t-T}^{t}
\delta_{\hat u}(\bx(\tau))\,d\tau.
\end{aligned}
\label{eq:integral_bellman_equation}
\end{equation}

From Eq.~\eqref{eq:critic_value_approximation} we have
$V^*(\bx(t))-V^*(\bx(t-T))
=
\bW^{*\top}
[\bphi(\bx(t))-\bphi(\bx(t-T))]
+
\varepsilon(\bx(t))
-
\varepsilon(\bx(t-T))$.
Define
$\Delta\bphi(t):=
\bphi(\bx(t))-\bphi(\bx(t-T))$.
Since
$\nabla V^{*\top}(\bx)
=
\bW^{*\top}\nabla\bphi(\bx)
+
\nabla\varepsilon^\top(\bx)$,
substitution into
\eqref{eq:integral_bellman_equation}
gives
\[
\begin{aligned}
0
={}&
\bW^{*\top}\Delta\bphi(t)
+
\varepsilon(\bx(t))
-
\varepsilon(\bx(t-T))
+
\int_{t-T}^{t}
\ell(\bx(\tau),\hat{\bu}(\tau))d\tau
\\
&-
\bW^{*\top}
\int_{t-T}^{t}
\nabla\bphi(\bx(\tau))
\bg(\bx(\tau))
[
\bu_b(\tau)-\hat{\bu}(\tau)
]d\tau
\\
&-
\int_{t-T}^{t}
\nabla\varepsilon^\top(\bx(\tau))
\bg(\bx(\tau))
[
\bu_b(\tau)-\hat{\bu}(\tau)
]d\tau
-
\int_{t-T}^{t}
\delta_{\hat u}(\bx(\tau))\,d\tau.
\end{aligned}
\]

Define
\begin{equation}
\begin{aligned}
\boldsymbol\Omega(t)
:={}&
\Delta\bphi(t)
-
\int_{t-T}^{t}
\nabla\bphi(\bx(\tau))
\bg(\bx(\tau))
\left[
\bu_b(\tau)-\hat{\bu}(\tau)
\right]d\tau.
\end{aligned}
\label{eq:integral_regressor}
\end{equation}
and
\begin{equation}
\rho(t)
:=
\int_{t-T}^{t}
\ell\!\left(
\bx(\tau),
\hat{\bu}(\tau)
\right)d\tau.
\label{eq:integral_cost_term}
\end{equation}

Define the aggregate Bellman residual as
\begin{equation}
\begin{aligned}
\varepsilon_B(t)
:={}&
\varepsilon(\bx(t))
-
\varepsilon(\bx(t-T))
\\
&-
\int_{t-T}^{t}
\nabla\varepsilon^\top(\bx(\tau))
\bg(\bx(\tau))
[
\bu_b(\tau)-\hat{\bu}(\tau)
]d\tau
\\
&-
\int_{t-T}^{t}
\delta_{\hat u}(\bx(\tau))\,d\tau.
\end{aligned}
\label{Residual}
\end{equation}

Using
\eqref{eq:integral_regressor}
and
\eqref{eq:integral_cost_term},
the integral Bellman equation becomes
\begin{equation}
    \bW^{*\top}\boldsymbol\Omega(t)
    +
    \rho(t)
    +
    \varepsilon_B(t)
    =
    0.
    \label{eq:integral_bellman_with_approximation}
\end{equation}


Equation~\eqref{eq:integral_bellman_with_approximation} is a significant equation in stability analysis; however, it is not the equation used for learning.
For the critic estimate \(\hat{\bW}\), define the integral Bellman
residual
\begin{equation}
    \xi(t)
    :=
    \hat{\bW}^{\top}(t)
    \boldsymbol\Omega(t)
    +
    \rho(t).
    \label{eq:critic_integral_residual}
\end{equation}

Motivated by the analysis in Proposition 1
of~\cite{Gong2025}, we have:

\begin{proposition}[Bounded Bellman approximation residual]
\label{pr2}
Under Proposition~\ref{prop:critic_approximation_residual},
suppose that the probing signal satisfies
$\|\be_p(t)\|\le e_M$.
If the learned policy is uniformly close
to the optimal policy over the considered
learning interval, then, for sufficiently
large $N_C$, there exists
$\mathscr K_B>0$ such that
$|\bar{\varepsilon}_B(t)|
\le\mathscr K_B$ and
$|\bar{\varepsilon}_{B,j}|
\le\mathscr K_B$,
$j=1,\ldots,M$.
\end{proposition}

\begin{proof}
By Proposition~\ref{prop:critic_approximation_residual},
for any $\eta>0$ and sufficiently large
$N_C$,
$|\varepsilon(\bx)|\le\eta$ and
$\|\nabla\varepsilon(\bx)\|
\le\bar\varepsilon_g\le\eta$.

From the optimal policy expression,
\begin{equation}
\begin{aligned}
\bu(\bx,\bW^*)-\bu^*(\bx)
=
\frac12\bR^{-1}\bg^\top(\bx)
\nabla\varepsilon(\bx),
\end{aligned}
\label{eq:ideal_policy_approximation_error}
\end{equation}
giving
$\|\bu(\bx,\bW^*)-\bu^*(\bx)\|
\le\frac12\|\bR^{-1}\|
G\bar\varepsilon_g$.
Hence, Proposition~\ref{prop:critic_approximation_residual}
ensures that the ideal-weight policy
approximation error can be made arbitrarily
small by increasing $N_C$, since the
gradient approximation error bound
$\bar\varepsilon_g$ can be reduced
accordingly as $N_C$ increases.
For the learned policy, under the stated
uniform policy-accuracy condition, let
$\|\hat{\bu}-\bu^*\|\le\hat\epsilon$.
Using the HJB stationary condition
$\partial H/\partial\bu
|_{\bu=\bu^*}
=\mathbf0$,
one has
$2\bR\bu^*
+\bm L_1^\top(\bx)
+\bg^\top(\bx)\nabla V^*(\bx)
=\mathbf0$.

Let
$\Delta\bu=\hat{\bu}-\bu^*$.
Then,
$\hat{\bu}=\bu^*+\Delta\bu$,
yielding
\begin{equation*}
\begin{aligned}
&\ell(\bx,\hat{\bu})
-\ell(\bx,\bu^*)
=
\bm L_1(\bx)\Delta\bu
+2\bu^{*\top}\bR\Delta\bu
+\Delta\bu^\top\bR\Delta\bu.
\end{aligned}
\end{equation*}

Consequently,
\begin{equation*}
\begin{aligned}
\delta_{\hat u}
={}&
\ell(\bx,\hat{\bu})
-\ell(\bx,\bu^*)
+\nabla V^{*\top}(\bx)\bg(\bx)
\Delta\bu
\\
={}&
\big(
\bm L_1^\top(\bx)
+2\bR\bu^*
+\bg^\top(\bx)\nabla V^*(\bx)
\big)^\top
\Delta\bu
\\
&+
\Delta\bu^\top\bR\Delta\bu.
\end{aligned}
\end{equation*}

By the HJB stationary condition and
$\bR\succ0$, it follows that
\begin{equation}
\begin{aligned}
0\le\delta_{\hat u}
&=
\Delta\bu^\top\bR\Delta\bu
\le
\lambda_{\max}(\bR)
\|\Delta\bu\|^2
\le
\lambda_{\max}(\bR)
\hat\epsilon^2.
\end{aligned}
\label{eq:policy_suboptimality_residual_bound}
\end{equation}

We have
$\bu_b-\hat{\bu}=\be_p$.
By the triangle and Cauchy--Schwarz
inequalities,
\begin{equation}
\begin{aligned}
|\varepsilon_B(t)|
&\le
2\eta+
\int_{t-T}^{t}
\|\nabla\varepsilon\|
\|\bg\|
\|\be_p\|\,d\tau
\\
&\quad+
\int_{t-T}^{t}
|\delta_{\hat u}|\,d\tau
\le
2\eta+
TG e_M\bar\varepsilon_g
+
T\lambda_{\max}(\bR)
\hat\epsilon^2
=:\mathscr K_B<\infty.
\end{aligned}
\label{eq:aggregate_residual_bound}
\end{equation}

The same bound applies to the historical
samples collected over the compact operating
set. Since the normalization denominators
are not smaller than one, the normalized
approximation residuals are also uniformly
bounded. Consequently, there exists a
finite constant $\mathscr K_B>0$
such that
$|\bar{\varepsilon}_B(t)|
\le\mathscr K_B$ and
$|\bar{\varepsilon}_{B,j}|
\le\mathscr K_B$,
for $j=1,\ldots,M$.
Moreover, increasing $N_C$ reduces
the $C^1$ approximation residual,
whereas the online policy error
additionally depends on critic learning.
\end{proof}

Using
\eqref{eq:integral_regressor}
and
\eqref{eq:integral_cost_term},
the integral Bellman equation becomes
\begin{equation}
\bW^{*\top}\boldsymbol\Omega(t)
+
\rho(t)
+
\varepsilon_B(t)
=
0.
\label{eq:integral_bellman_with_approximation}
\end{equation}

Equation~\eqref{eq:integral_bellman_with_approximation} is a significant equation in stability analysis; however, it is not the equation used for learning.
For the critic estimate \(\hat{\bW}\), define the integral Bellman
residual
\begin{equation}
    \xi(t)
    :=
    \hat{\bW}^{\top}(t)
    \boldsymbol\Omega(t)
    +
    \rho(t).
    \label{eq:critic_integral_residual}
\end{equation}

Let
\(\tilde{\bW}:=
\hat{\bW}-\bW^*\).
Under exact approximation,
\(\bW^{*\top}\boldsymbol\Omega+\rho=0\),
and hence
\(\xi=
\hat{\bW}^{\top}\boldsymbol\Omega
-
\bW^{*\top}\boldsymbol\Omega
=
\tilde{\bW}^{\top}\boldsymbol\Omega\).
More generally, when
\(\varepsilon\neq0\),
\eqref{eq:integral_bellman_with_approximation} gives
\(\rho=
-\bW^{*\top}\boldsymbol\Omega
-
\varepsilon_B\),
and therefore
\(\xi=
\hat{\bW}^{\top}\boldsymbol\Omega
-
\bW^{*\top}\boldsymbol\Omega
-
\varepsilon_B
=
\tilde{\bW}^{\top}\boldsymbol\Omega
-
\varepsilon_B\).

To normalize the regressor, define
\begin{equation}
    m(t)
    :=
    1+
    \boldsymbol\Omega^\top(t)
    \boldsymbol\Omega(t),
    \qquad
    \bar{\boldsymbol\Omega}(t)
    :=
    \frac{\boldsymbol\Omega(t)}
    {m(t)} .
    \label{eq:normalized_integral_regressor}
\end{equation}
The normalized Bellman residual is
\begin{equation}
    e(t)
    :=
    \frac{\xi(t)}{m(t)}
    =
    \frac{
        \hat{\bW}^{\top}(t)\boldsymbol\Omega(t)
        +
        \rho(t)
    }{
        m(t)
    } .
    \label{eq:normalized_bellman_error}
\end{equation}
Under exact approximation,
\(e(t)=
\tilde{\bW}^{\top}
\bar{\boldsymbol\Omega}(t)\).
For the general approximation case, define
\(\bar\varepsilon_B(t):=
\varepsilon_B(t)/m(t)\).
Then
\(e(t)=
\tilde{\bW}^{\top}
\bar{\boldsymbol\Omega}(t)
-
\bar\varepsilon_B(t)\),
or equivalently,
\(\tilde{\bW}^{\top}
\bar{\boldsymbol\Omega}(t)
=
e(t)
+
\bar\varepsilon_B(t)\).

Let \(M\) previously recorded integral samples be denoted by
\(\{\boldsymbol\Omega_j,\rho_j\}_{j=1}^{M}\).
For each stored sample, define
\(m_j:=
1+\boldsymbol\Omega_j^\top\boldsymbol\Omega_j\),
\(\bar{\boldsymbol\Omega}_j:=
\boldsymbol\Omega_j/m_j\),
and
\begin{equation}
    e_j
    :=
    \frac{
        \hat{\bW}^{\top}\boldsymbol\Omega_j
        +
        \rho_j
    }{
        m_j
    } .
    \label{eq:recorded_bellman_error}
\end{equation}

For each recorded sample, define
\(\varepsilon_{B,j}\)
according to
\eqref{Residual}
evaluated on the corresponding stored window and
\(\bar\varepsilon_{B,j}:=
\varepsilon_{B,j}/m_j\).
Then
\(e_j=
\tilde{\bW}^{\top}
\bar{\boldsymbol\Omega}_j
-
\bar\varepsilon_{B,j}\),
or equivalently,
\(\tilde{\bW}^{\top}
\bar{\boldsymbol\Omega}_j
=
e_j
+
\bar\varepsilon_{B,j}\).


The recorded regressors are assumed to satisfy
\begin{equation}
    \sum_{j=1}^{M}
    \bar{\boldsymbol\Omega}_j
    \bar{\boldsymbol\Omega}_j^\top
    \succeq
    \alpha_{\mathrm{IE}}\bm I,
    \qquad
    \alpha_{\mathrm{IE}}>0 .
    \label{eq:critic_ie_condition}
\end{equation}
Equivalently,
\(\lambda_{\min}
(
\sum_{j=1}^{M}
\bar{\boldsymbol\Omega}_j
\bar{\boldsymbol\Omega}_j^\top
)
\geq
\alpha_{\mathrm{IE}}>0\).

Define
\(\operatorname{sig}^{\gamma}(s)
:=
|s|^\gamma\operatorname{sgn}(s)\).
Consider the scalar two-power penalty
\(\Psi(e)
:=
\frac{\eta_{W1}}{\gamma_1+1}|e|^{\gamma_1+1}
+
\frac{\eta_{W2}}{\gamma_2+1}|e|^{\gamma_2+1}\),
where \(\eta_{W1},\eta_{W2}>0\). Its derivative is
\(\frac{\partial\Psi(e)}{\partial e}
=
\eta_{W1}\operatorname{sig}^{\gamma_1}(e)
+
\eta_{W2}\operatorname{sig}^{\gamma_2}(e)\).

Since
\(e(t)=
[
\hat{\bW}^{\top}\boldsymbol\Omega(t)+\rho(t)
]/m(t)\)
and by chain rule,
one has
\(\partial e(t)/\partial\hat{\bW}
=
\boldsymbol\Omega(t)/m(t)
=
\bar{\boldsymbol\Omega}(t)\).
Therefore,
\(\nabla_{\hat{\bW}}\Psi(e(t))
=
\bar{\boldsymbol\Omega}(t)
[
\eta_{W1}\operatorname{sig}^{\gamma_1}(e(t))
+
\eta_{W2}\operatorname{sig}^{\gamma_2}(e(t))
]\).

Similarly,
\(\partial e_j/\partial\hat{\bW}
=
\bar{\boldsymbol\Omega}_j\),
and hence
\(\nabla_{\hat{\bW}}\Psi(e_j)
=
\bar{\boldsymbol\Omega}_j
[
\eta_{W1}\operatorname{sig}^{\gamma_1}(e_j)
+
\eta_{W2}\operatorname{sig}^{\gamma_2}(e_j)
]\).

Define the composite critic objective
\begin{equation}
    \mathcal J_W
    :=
    \Psi(e(t))
    +
    \sum_{j=1}^{M}
    \Psi(e_j).
    \label{eq:critic_composite_objective}
\end{equation}
Then
\[
\begin{aligned}
\nabla_{\hat{\bW}}\mathcal J_W
={}&
\bar{\boldsymbol\Omega}(t)
\left[
\eta_{W1}\operatorname{sig}^{\gamma_1}(e(t))
+
\eta_{W2}\operatorname{sig}^{\gamma_2}(e(t))
\right]
\\
&+
\sum_{j=1}^{M}
\bar{\boldsymbol\Omega}_j
\left[
\eta_{W1}\operatorname{sig}^{\gamma_1}(e_j)
+
\eta_{W2}\operatorname{sig}^{\gamma_2}(e_j)
\right].
\end{aligned}
\]

To introduce an explicit scaling parameter controlling the resulting
fixed-time critic-learning bound,
consider the scaled gradient flow
\(
\dot{\hat{\bW}}
=
-\alpha_W
\bm{\Gamma}
\nabla_{\hat{\bW}}\mathcal J_W
\),
where
\(\alpha_W>0\) and
\(\bm{\Gamma}
=
\bm{\Gamma}^{\top}
\succ0\).
This gives the fixed-time critic update law

\begin{equation}
\begin{aligned}
    \dot{\hat{\bW}}
    ={}&
    -\alpha_W
    \bm{\Gamma}
    \bar{\boldsymbol\Omega}(t)
    \left[
        \eta_{W1}
        \operatorname{sig}^{\gamma_1}(e(t))
        +
        \eta_{W2}
        \operatorname{sig}^{\gamma_2}(e(t))
    \right]
    \\
    &-
    \alpha_W
    \bm{\Gamma}
    \sum_{j=1}^{M}
    \bar{\boldsymbol\Omega}_j
    \left[
        \eta_{W1}
        \operatorname{sig}^{\gamma_1}(e_j)
        +
        \eta_{W2}
        \operatorname{sig}^{\gamma_2}(e_j)
    \right].
\end{aligned}
\label{eq:fixed_time_critic_update}
\end{equation}

\begin{remark}[Distinction from existing learning schemes]
~\cite{Wang2024FixedTime} investigated
fixed-time reinforcement learning but did not employ
experience replay. In contrast,
~\cite{Gong2025} and
~\cite{Kokolakis2025} incorporated experience-replay
mechanisms. However, ~\cite{Gong2025} did not explicitly
construct a running cost designed to induce
fixed-time stability, and their learning law was
developed under the assumption of known system
dynamics. Similarly, ~\cite{Kokolakis2025} did not
explicitly clarify the construction of the running
cost and its connection to the derivation of the
critic update law, while also considering known
system dynamics. ~\cite{Wang2026SafeFixedTime}
further investigated safe fixed-time reinforcement
learning with experience replay in a zero-sum
game setting. The fundamental distinction of the
present work lies in the integral Bellman residual
$e(t)$, defined in~\eqref{eq:normalized_integral_regressor},
which is constructed from off-policy data over
finite trajectory windows. Consequently, the
proposed critic update
\eqref{eq:fixed_time_critic_update}
is fundamentally different in its underlying
regression formulation. Nevertheless, the
two-power non-Lipschitz structure follows
a similar principle of
~\cite{Gong2025,Kokolakis2025Predefined,Wang2026SafeFixedTime}
for achieving fixed-time or predefined-time
convergence. Furthermore, the present work
explicitly constructs an inverse-optimal running
cost consistent with the prescribed Lyapunov decay
while identifying the unknown drift through ICL.
These distinctions are further clarified by the
stability proofs of
Theorems~\ref{thm:fixed_time_critic}
and~\ref{thm:closed_loop}.
\end{remark}

\begin{lemma}[Replay-stack coercivity]
\label{lem:replay_stack_coercivity}
Suppose that the recorded normalized regressors satisfy the finite-data
informativity condition \eqref{eq:critic_ie_condition}, namely,
\(
\sum_{j=1}^{M}
\bar{\boldsymbol\Omega}_j
\bar{\boldsymbol\Omega}_j^\top
\succeq
\alpha_{\mathrm{IE}}\bm I
\),
with
\(
\alpha_{\mathrm{IE}}>0
\).
Then, for any
\(
\tilde{\bW}\in\mathbb R^{N_c}
\),
the following inequalities hold:
\begin{equation}
\sum_{j=1}^{M}
\left|
\tilde{\bW}^{\top}
\bar{\boldsymbol\Omega}_j
\right|^{\gamma_1+1}
\geq
\alpha_{\mathrm{IE}}^{\frac{\gamma_1+1}{2}}
\|\tilde{\bW}\|^{\gamma_1+1},
\label{eq:replay_lower_bound_gamma1}
\end{equation}
and
\begin{equation}
\sum_{j=1}^{M}
\left|
\tilde{\bW}^{\top}
\bar{\boldsymbol\Omega}_j
\right|^{\gamma_2+1}
\geq
M^{\frac{1-\gamma_2}{2}}
\alpha_{\mathrm{IE}}^{\frac{\gamma_2+1}{2}}
\|\tilde{\bW}\|^{\gamma_2+1}.
\label{eq:replay_lower_bound_gamma2}
\end{equation}
In particular, both lower-bound coefficients are strictly positive.
\end{lemma}

\begin{proof}
Define the replay-regressor matrix
\(
\bar{\bm\Omega}
:=
[
\bar{\boldsymbol\Omega}_1,\ldots,
\bar{\boldsymbol\Omega}_M
]
\).
Then
\(
\bar{\bm\Omega}^{\top}\tilde{\bW}
=
\operatorname{col}
\{
\tilde{\bW}^{\top}\bar{\boldsymbol\Omega}_1,
\ldots,
\tilde{\bW}^{\top}\bar{\boldsymbol\Omega}_M
\}
\).
From \eqref{eq:critic_ie_condition},
\(
\bar{\bm\Omega}\bar{\bm\Omega}^{\top}
\succeq
\alpha_{\mathrm{IE}}\bm I
\),
and therefore
\(
\sigma_{\min}(\bar{\bm\Omega})
\geq
\sqrt{\alpha_{\mathrm{IE}}}
\).
Hence,
\(
\|\bar{\bm\Omega}^{\top}\tilde{\bW}\|_2
\geq
\sqrt{\alpha_{\mathrm{IE}}}\,
\|\tilde{\bW}\|
\).

Since
\(
0<\gamma_1<1
\),
one has
\(
1<\gamma_1+1<2
\).
For any finite-dimensional vector, the
\(
\ell_p
\)-norm is no smaller than the
\(
\ell_2
\)-norm when
\(
1<p<2
\).
Thus,
\[
\begin{aligned}
\sum_{j=1}^{M}
\left|
\tilde{\bW}^{\top}
\bar{\boldsymbol\Omega}_j
\right|^{\gamma_1+1}
&=
\left\|
\bar{\bm\Omega}^{\top}\tilde{\bW}
\right\|_{\gamma_1+1}^{\gamma_1+1}
\geq
\left\|
\bar{\bm\Omega}^{\top}\tilde{\bW}
\right\|_2^{\gamma_1+1}
\geq
\alpha_{\mathrm{IE}}^{\frac{\gamma_1+1}{2}}
\|\tilde{\bW}\|^{\gamma_1+1},
\end{aligned}
\]
which proves \eqref{eq:replay_lower_bound_gamma1}.

Since
\(
\gamma_2>1
\),
one has
\(
\gamma_2+1>2
\).
Using the finite-dimensional norm relation
\(
\|z\|_p
\geq
M^{\frac{1}{p}-\frac{1}{2}}
\|z\|_2
\)
for
\(
p>2
\),
with
\(
p=\gamma_2+1
\),
gives
\[
\left\|
\bar{\bm\Omega}^{\top}\tilde{\bW}
\right\|_{\gamma_2+1}
\geq
M^{
\frac{1}{\gamma_2+1}
-
\frac{1}{2}
}
\left\|
\bar{\bm\Omega}^{\top}\tilde{\bW}
\right\|_2.
\]
Raising both sides to the power
\(
\gamma_2+1
\)
and using the finite-data lower bound yields
\[
\begin{aligned}
\sum_{j=1}^{M}
\left|
\tilde{\bW}^{\top}
\bar{\boldsymbol\Omega}_j
\right|^{\gamma_2+1}
&=
\left\|
\bar{\bm\Omega}^{\top}\tilde{\bW}
\right\|_{\gamma_2+1}^{\gamma_2+1}
\geq
M^{\frac{1-\gamma_2}{2}}
\left\|
\bar{\bm\Omega}^{\top}\tilde{\bW}
\right\|_2^{\gamma_2+1}
\geq
M^{\frac{1-\gamma_2}{2}}
\alpha_{\mathrm{IE}}^{\frac{\gamma_2+1}{2}}
\|\tilde{\bW}\|^{\gamma_2+1},
\end{aligned}
\]
which proves \eqref{eq:replay_lower_bound_gamma2}. Since
\(
M\geq1
\)
and
\(
\alpha_{\mathrm{IE}}>0
\),
both coefficients are strictly positive.
\end{proof}

\begin{lemma}[Perturbed signed-power inequality]
\label{lem:perturbed_power}
For every \(\gamma>0\) and all \(z,\epsilon\in\mathbb R\),
\begin{equation}
z\spow{-z+\epsilon}{\gamma}
\le
-2^{-(\gamma+1)}
|z|^{\gamma+1}
+
\left(
1+2\,3^\gamma
\right)
|\epsilon|^{\gamma+1}.
\label{eq:perturbed_power}
\end{equation}
If \(\epsilon=0\), the sharper identity
\begin{equation}
z\spow{-z}{\gamma}
=
-|z|^{\gamma+1}
\label{eq:exact_signed_power_identity}
\end{equation}
holds.
\end{lemma}

\begin{theorem}[Fixed-time critic identification from finite replay data]
\label{thm:fixed_time_critic}
Consider the critic adaptation law
\eqref{eq:fixed_time_critic_update} with
\(0<\gamma_1<1<\gamma_2\), \(\eta_{W1},\eta_{W2},\alpha_W>0\),
and \(\bm{\Gamma}\succ0\). Suppose that the recorded regressors
satisfy the finite-data informativity condition
\eqref{eq:critic_ie_condition}, and that
the informative replay stack remains available
throughout the critic-learning process.
The time origin is taken at the instant
when this condition is satisfied.
For the critic-weight error
\(\tilde{\bW}=\hat{\bW}-\bW^*\), let
\begin{equation}
\begin{aligned}
V_W&:=\frac12\tilde{\bW}^{\top}
\bm{\Gamma}^{-1}\tilde{\bW},
a_W:=\eta_{W1}\left(2\lambda_{\min}(\bm{\Gamma})
\alpha_{\mathrm{IE}}\right)^\mu,\\
b_W&:=\eta_{W2}M^{\frac{1-\gamma_2}{2}}
\left(2\lambda_{\min}(\bm{\Gamma})
\alpha_{\mathrm{IE}}\right)^\nu.
\end{aligned}
\label{eq:critic_fixed_coefficients}
\end{equation}
Then the following fixed-time convergence properties hold,
independently of the initial critic-weight error.

\emph{(i) Exact Bellman data:}
If \(\bar\varepsilon_B(t)=0\) and
\(\bar\varepsilon_{B,j}=0\) for \(j=1,\ldots,M\),
the critic weights recover the ideal weights in fixed time:
\begin{equation}
\tilde{\bW}(t)=\bm0,\qquad
t\geq T_{W,0},
\label{eq:critic_exact_fixed_convergence}
\end{equation}
where
\begin{equation}
T_{W,0}:=
\frac{1}{\alpha_Wa_W(1-\mu)}
+\frac{1}{\alpha_Wb_W(\nu-1)}.
\label{eq:critic_exact_time_bound}
\end{equation}

\emph{(ii) Perturbed Bellman data:}
Following Proposition~\ref{pr2}, we define
For any fixed \(\chi_W\in(0,1)\), define
\begin{equation}
a_{W\Delta}:=2^{-(\gamma_1+1)}a_W,
\qquad
b_{W\Delta}:=2^{-(\gamma_2+1)}b_W,
\label{eq:perturbed_critic_coefficients}
\end{equation}
and let \(\Delta_W>0\) be the aggregate perturbation
specified in \eqref{eq:critic_residual_perturbation}.
Let \(\bar V_W\) be the unique positive solution of
\begin{equation}
\alpha_W(1-\chi_W)
\left(a_{W\Delta}\bar V_W^\mu
+b_{W\Delta}\bar V_W^\nu\right)=\Delta_W.
\label{eq:critic_fixed_residual_radius}
\end{equation}
Then the residual set
\begin{equation}
\mathcal B_W:=\left\{V_W\in\mathbb R_{\geq0}:
V_W\leq\bar V_W\right\}
\label{eq:critic_fixed_invariant_set}
\end{equation}
is forward invariant and reached no later than
\begin{equation}
T_{W,\Delta}:=
\frac{1}{\alpha_W\chi_Wa_{W\Delta}(1-\mu)}
+\frac{1}{\alpha_W\chi_Wb_{W\Delta}(\nu-1)}.
\label{eq:critic_practical_time_bound}
\end{equation}
In particular,
\begin{equation}
\|\tilde{\bW}(t)\|
\leq\sqrt{2\lambda_{\max}(\bm{\Gamma})\bar V_W},
\qquad t\geq T_{W,\Delta}.
\label{eq:critic_fixed_weight_bound}
\end{equation}
\end{theorem}

\begin{proof}
Since \(\bW^*\) is constant and
\(\tilde{\bW}=\hat{\bW}-\bW^*\), one has
\(\dot{\tilde{\bW}}=\dot{\hat{\bW}}\).
Consider the critic Lyapunov function
\(V_W\) defined in \eqref{eq:critic_fixed_coefficients}.
Differentiating \(V_W\) and substituting
\eqref{eq:fixed_time_critic_update} yield
\begin{equation}
\begin{aligned}
\dot V_W
={}&
-\alpha_W
\tilde{\bW}^{\top}
\bar{\boldsymbol\Omega}(t)
\left[
\eta_{W1}\spow{e(t)}{\gamma_1}
+
\eta_{W2}\spow{e(t)}{\gamma_2}
\right]
\\
&-
\alpha_W
\sum_{j=1}^{M}
\tilde{\bW}^{\top}
\bar{\boldsymbol\Omega}_j
\left[
\eta_{W1}\spow{e_j}{\gamma_1}
+
\eta_{W2}\spow{e_j}{\gamma_2}
\right].
\end{aligned}
\label{eq:VW_dot_raw}
\end{equation}

For the exact-approximation case,
\(\bar\varepsilon_B(t)=0\) and
\(\bar\varepsilon_{B,j}=0\), and hence
\(e(t)=\tilde{\bW}^{\top}\bar{\boldsymbol\Omega}(t)\) and
\(e_j=\tilde{\bW}^{\top}\bar{\boldsymbol\Omega}_j\).
Using the identity
\(z\spow{z}{\gamma}=|z|^{\gamma+1}\)
in \eqref{eq:VW_dot_raw} gives
\begin{equation}
\begin{aligned}
\dot V_W
={}&
-\alpha_W \eta_{W1}
\left|
\tilde{\bW}^{\top}
\bar{\boldsymbol\Omega}(t)
\right|^{\gamma_1+1}
-
\alpha_W \eta_{W2}
\left|
\tilde{\bW}^{\top}
\bar{\boldsymbol\Omega}(t)
\right|^{\gamma_2+1}
\\
&-
\alpha_W \eta_{W1}
\sum_{j=1}^{M}
\left|
\tilde{\bW}^{\top}
\bar{\boldsymbol\Omega}_j
\right|^{\gamma_1+1}
-
\alpha_W \eta_{W2}
\sum_{j=1}^{M}
\left|
\tilde{\bW}^{\top}
\bar{\boldsymbol\Omega}_j
\right|^{\gamma_2+1}.
\end{aligned}
\label{eq:VW_exact_raw}
\end{equation}

Discarding the nonpositive current-data terms and applying
Lemma~\ref{lem:replay_stack_coercivity} yield
\begin{equation}
\begin{aligned}
\dot V_W
\leq{}&
-\alpha_W \eta_{W1}
\alpha_{\mathrm{IE}}^{\frac{\gamma_1+1}{2}}
\|\tilde{\bW}\|^{\gamma_1+1}
\\
&-
\alpha_W \eta_{W2}
M^{\frac{1-\gamma_2}{2}}
\alpha_{\mathrm{IE}}^{\frac{\gamma_2+1}{2}}
\|\tilde{\bW}\|^{\gamma_2+1}.
\end{aligned}
\label{eq:VW_exact_weight}
\end{equation}

From the definition of \(V_W\),
\(
V_W
\leq
\|\tilde{\bW}\|^2/
[2\lambda_{\min}(\bm{\Gamma})]
\),
and therefore
\(
\|\tilde{\bW}\|^2
\geq
2\lambda_{\min}(\bm{\Gamma})V_W
\).
Using this relation in
\eqref{eq:VW_exact_weight} and recalling
\(\mu=(\gamma_1+1)/2\) and
\(\nu=(\gamma_2+1)/2\) gives
\begin{equation}
\dot V_W
\leq
-\alpha_W
\left[
a_WV_W^\mu
+
b_WV_W^\nu
\right].
\label{eq:VW_exact_comparison}
\end{equation}

Since \(\alpha_W>0\), applying
Theorem~\ref{thm:fixed_time_stability} to
\eqref{eq:VW_exact_comparison} yields
\(T_W^{\mathrm{act}}\leq T_{W,0}\), independently of
\(V_W(0)\), where \(T_W^{\mathrm{act}}\) is the actual
convergence time. Consequently,
\[
\tilde{\bW}(t)=\bm0,
\qquad
t\geq T_{W,0},
\]
which proves the exact-approximation claim.

For the nonzero-residual case,
\[
e(t)
=
\tilde{\bW}^{\top}
\bar{\boldsymbol\Omega}(t)
-
\bar\varepsilon_B(t),
\qquad
e_j
=
\tilde{\bW}^{\top}
\bar{\boldsymbol\Omega}_j
-
\bar\varepsilon_{B,j}.
\]

Applying Lemma~\ref{lem:perturbed_power} with
\(\gamma=\gamma_1\) and \(\gamma=\gamma_2\) to every term in
\eqref{eq:VW_dot_raw}, using
Proposition~\ref{prop:critic_approximation_residual}, and discarding the
remaining nonpositive current-data dissipation give
\begin{equation}
\begin{aligned}
\dot V_W
\leq{}&
-\alpha_W \eta_{W1}
2^{-(\gamma_1+1)}
\sum_{j=1}^{M}
\left|
\tilde{\bW}^{\top}
\bar{\boldsymbol\Omega}_j
\right|^{\gamma_1+1}
\\
&-
\alpha_W \eta_{W2}
2^{-(\gamma_2+1)}
\sum_{j=1}^{M}
\left|
\tilde{\bW}^{\top}
\bar{\boldsymbol\Omega}_j
\right|^{\gamma_2+1}
+
\Delta_W,
\end{aligned}
\label{eq:VW_perturbed_before_coercivity}
\end{equation}
where
\begin{equation}
\begin{aligned}
\Delta_W
:={}&
\alpha_W(M+1)
\Big[
\eta_{W1}
\left(
1+2\,3^{\gamma_1}
\right)
\mathscr K_B^{\gamma_1+1}
\\
&\qquad\qquad
+
\eta_{W2}
\left(
1+2\,3^{\gamma_2}
\right)
\mathscr K_B^{\gamma_2+1}
\Big].
\end{aligned}
\label{eq:critic_residual_perturbation}
\end{equation}

Lemma~\ref{lem:replay_stack_coercivity} and
\(
\|\tilde{\bW}\|^2
\geq
2\lambda_{\min}(\bm{\Gamma})V_W
\)
now imply
\begin{equation}
\dot V_W
\leq
-\alpha_W
\left[
a_{W\Delta}V_W^\mu
+
b_{W\Delta}V_W^\nu
\right]
+
\Delta_W.
\label{eq:VW_practical}
\end{equation}

By the definition of \(\bar V_W\) in
\eqref{eq:critic_fixed_residual_radius}, for every
\(V_W>\bar V_W\), the strict monotonicity of the two power terms gives
\[
\Delta_W
<
\alpha_W(1-\chi_W)
\left[
a_{W\Delta}V_W^\mu
+
b_{W\Delta}V_W^\nu
\right].
\]
Substituting this inequality into \eqref{eq:VW_practical} yields,
for every \(V_W\notin\mathcal B_W\),
\begin{equation}
\dot V_W
\leq
-\alpha_W\chi_W
\left[
a_{W\Delta}V_W^\mu
+
b_{W\Delta}V_W^\nu
\right].
\label{eq:VW_outside_BW}
\end{equation}

Applying the fixed-time comparison argument of
Theorem~\ref{thm:fixed_time_stability} up to the first entry
into \(\mathcal B_W\) shows that the entrance time into
\(\mathcal B_W\) satisfies
\[
T_{W,\mathcal B_W}\leq T_{W,\Delta},
\]
where \(T_{W,\mathcal B_W}\) is the time of entry into
the residual set \(\mathcal B_W\).
Consequently,
\begin{equation}
V_W(t)
\leq
\bar V_W,
\qquad
t\geq T_{W,\Delta}.
\label{eq:VW_practical_entry}
\end{equation}

It remains to verify forward invariance. At the boundary
\(V_W=\bar V_W\), equations
\eqref{eq:VW_practical} and
\eqref{eq:critic_fixed_residual_radius} yield
\begin{equation}
\begin{aligned}
\dot V_W
&\leq
-\alpha_W
\left[
a_{W\Delta}\bar V_W^\mu
+
b_{W\Delta}\bar V_W^\nu
\right]
+
\Delta_W
\\
&=
-\alpha_W\chi_W
\left[
a_{W\Delta}\bar V_W^\mu
+
b_{W\Delta}\bar V_W^\nu
\right]
\leq 0.
\end{aligned}
\label{eq:VW_boundary_invariance}
\end{equation}
Therefore, \(\mathcal B_W\) is forward invariant.

Finally, from
\[
V_W
=
\frac{1}{2}
\tilde{\bW}^{\top}
\bm{\Gamma}^{-1}
\tilde{\bW},
\]
one has
\[
V_W
\geq
\frac{
\|\tilde{\bW}\|^2
}{
2\lambda_{\max}(\bm{\Gamma})
}.
\]
Combining this relation with \eqref{eq:VW_practical_entry} gives
\[
\|\tilde{\bW}(t)\|
\leq
\sqrt{
2\lambda_{\max}(\bm{\Gamma})
\bar V_W
},
\qquad
t\geq T_{W,\Delta}.
\]
This proves the practical fixed-time critic-convergence claim.
\end{proof}

\section{Fixed-Time Closed-Loop Stability Analysis}
\label{sec:closed_loop}

Define the composite quantity
\begin{equation}
\begin{aligned}
\mathcal J(t)
&:=
\mathcal I_x(t)+V_W(t),\quad
\mathcal I_x(t)
:=
\int_{t-T}^{t}
V_d(\bx(\tau))\,d\tau.
\end{aligned}
\label{eq:closed_loop_J}
\end{equation}

Here, \(V_W\) is the critic-weight Lyapunov
function defined in
Theorem~\ref{thm:fixed_time_critic}.
The use of \(V_d\) in \(\mathcal I_x\)
permits direct application of the desired
fixed-time evolution
\eqref{eq:desired_fixed_time_decay},
without requiring the plant-optimal value
function \(V^*\) to coincide with \(V_d\).

Since the closed-loop analysis is carried
out on the compact set \(\Omega\), and
\(V_d\), \(V^*\), \(\bg\), \(\bphi\), and
\(\varepsilon\) are continuously differentiable
on \(\Omega\), the corresponding quantities
are uniformly bounded. Define
\(
r_\Omega
:=
\sup_{\bm\xi\in\Omega}\|\bm\xi\|
\),
\(
\bar p_d
:=
\sup_{\bm\xi\in\Omega}
\|\nabla V_d(\bm\xi)\|
\),
\(
\bar g
:=
\sup_{\bm\xi\in\Omega}
\|\bg(\bm\xi)\|
\),
\(
\bar\phi_g
:=
\sup_{\bm\xi\in\Omega}
\|\nabla\bphi(\bm\xi)\|
\),
and
\(
\bar\varepsilon_g
:=
\sup_{\bm\xi\in\Omega}
\|\nabla\varepsilon(\bm\xi)\|
\).

From
\(
V^*
=
\bW^{*\top}\bphi+\varepsilon
\)
and
\(
\hat V
=
\hat{\bW}^{\top}\bphi
\),
one has
\(
\nabla V^*-\nabla\hat V
=
-(\nabla\bphi)^\top\tilde{\bW}
+
\nabla\varepsilon
\),
and therefore
\(
\|\nabla V^*-\nabla\hat V\|
\leq
\bar\phi_g\|\tilde{\bW}\|
+
\bar\varepsilon_g
\).

Moreover, comparing the critic policy
\eqref{eq:critic_only_policy}
with the inverse-optimal minimum-effort
policy
\eqref{eq:continuous_fixed_time_controller}
yields
\(
\hat{\bu}-\bu_c
=
-\frac12
\bR^{-1}\bg^\top
[\nabla\hat V-\nabla V_d]
\).
Consequently,
\(
\|\nabla\hat V-\nabla V_d\|
\leq
\bar\phi_g\|\tilde{\bW}\|
+
\bar\varepsilon_g+\bar d_V
\),
and
\(
\|\hat{\bu}-\bu_c\|
\leq
\frac12
\|\bR^{-1}\|\bar g
[
\bar\phi_g\|\tilde{\bW}\|
+
\bar\varepsilon_g+\bar d_V
]
\).

Thus,
\(
|
\nabla V_d^\top
\bg(\hat{\bu}-\bu_c)
|
\leq
c_W\|\tilde{\bW}\|+c_0
\),
where
\begin{equation}
\begin{aligned}
c_W
&:=
\frac12
\bar p_d
\|\bR^{-1}\|
\bar g^2
\bar\phi_g,\quad
c_0
:=
\frac12
\bar p_d
\|\bR^{-1}\|
\bar g^2
(\bar\varepsilon_g+\bar d_V).
\end{aligned}
\label{eq:closed_loop_mismatch_constants}
\end{equation}

\begin{remark}
Equation~\eqref{eq:desired_fixed_time_decay}
has already been established for \(V_d\)
under the inverse-optimal feedback
\(\bu_c\) along the identified model.
The critic-generated policy \(\hat{\bu}\)
is generally different from \(\bu_c\),
and the actual drift \(\bfv\) differs
from \(\hat{\bfv}\).
Accordingly, these discrepancies are
treated explicitly as perturbations.
\end{remark}

For the actual closed-loop dynamics
\(
\dot{\bx}
=
\bfv(\bx)+\bg(\bx)\hat{\bu}
\),
write
\(
\bfv(\bx)
=
\hat{\bfv}(\bx)
+
\tilde{\bfv}(\bx)
\),
where
\(
\tilde{\bfv}
:=
\bfv-\hat{\bfv}
\).
Then
\begin{equation}
\begin{aligned}
\dot V_d
={}&
\nabla V_d^\top
[
\hat{\bfv}+\bg\bu_c
]
+
\nabla V_d^\top\tilde{\bfv}
+
\nabla V_d^\top\bg
(\hat{\bu}-\bu_c).
\end{aligned}
\label{eq:Vstar_actual_decomposition}
\end{equation}

By the inverse-optimal construction and
\eqref{eq:desired_fixed_time_decay},
\(
\nabla V_d^\top
[
\hat{\bfv}+\bg\bu_c
]
=
-k_1V_d^{\gamma_1}
-k_2V_d^\nu
\).

By Lemma~\ref{lem:identification_error},
\(
\|\tilde{\bfv}(\bx)\|
\leq
\bar\Delta_f
\),
and hence
\(
\nabla V_d^\top\tilde{\bfv}
\leq
\bar p_d\bar\Delta_f
\).
Define
\(
c_f
:=
\bar p_d\bar\Delta_f
\).

Using the policy-error bound, one obtains
\begin{equation}
\dot V_d
\leq
-k_1V_d^{\gamma_1}
-k_2V_d^\nu
+
c_W\|\tilde{\bW}\|
+
c_0+c_f.
\label{eq:Vstar_learned}
\end{equation}

\begin{lemma}[Optimized Young inequality~\citep{Hardy1988Inequalities}]
\label{lem:optimized_young}
For
\(s\geq0\),
\(c\geq0\),
\(b>0\),
and
\(p>1\),
\begin{equation}
cs
\leq
bs^p
+
(p-1)
p^{-\frac{p}{p-1}}
b^{-\frac1{p-1}}
c^{\frac p{p-1}}.
\label{eq:optimized_Young}
\end{equation}
\end{lemma}

For arbitrary
\(b_1,b_2>0\),
write
\(
c_W\|\tilde{\bW}\|
=
(c_W/2)\|\tilde{\bW}\|
+
(c_W/2)\|\tilde{\bW}\|
\).
Applying
Lemma~\ref{lem:optimized_young}
to the two terms with
\(p=\gamma_1+1\)
and
\(p=\gamma_2+1\),
respectively, yields
\(
c_W\|\tilde{\bW}\|
+
c_0+c_f
\leq
b_1\|\tilde{\bW}\|^{\gamma_1+1}
+
b_2\|\tilde{\bW}\|^{\gamma_2+1}
+
\Delta_x
\),
where
\begin{equation}
\begin{aligned}
\Delta_x
:={}&
c_0+c_f
+
\gamma_1
(\gamma_1+1)^{
-\frac{\gamma_1+1}{\gamma_1}}
b_1^{-\frac1{\gamma_1}}
\left(
\frac{c_W}{2}
\right)^{
\frac{\gamma_1+1}{\gamma_1}}
\\
&+
\gamma_2
(\gamma_2+1)^{
-\frac{\gamma_2+1}{\gamma_2}}
b_2^{-\frac1{\gamma_2}}
\left(
\frac{c_W}{2}
\right)^{
\frac{\gamma_2+1}{\gamma_2}}.
\end{aligned}
\label{eq:Delta_x}
\end{equation}

Thus,
\begin{equation}
\begin{aligned}
\dot V_d
\leq{}&
-k_1V_d^{\gamma_1}
-k_2V_d^\nu
+
b_1\|\tilde{\bW}\|^{\gamma_1+1}
\\
&+
b_2\|\tilde{\bW}\|^{\gamma_2+1}
+
\Delta_x.
\end{aligned}
\label{eq:Vstar_learned_bound}
\end{equation}

Since \(V_d\) is continuous on the compact
set \(\Omega\), it is uniformly bounded
on \(\Omega\). Define
\(
\bar V_d
:=
\sup_{\bm\xi\in\Omega}V_d(\bm\xi)
\)
and
\(
\bar I_x
:=
T\bar V_d
\),
so that
\(
0\leq\mathcal I_x(t)\leq\bar I_x
\).
we also have
\(
|\dot V_d(\bx(t))|
\leq
L_{\dot V}
\)
for some finite
\(L_{\dot V}>0\)
along the considered closed-loop trajectories.

\begin{lemma}[Derivative-limited low-order window bound]
\label{lem:low_window}
Let
\(y:[t-T,t]\rightarrow\mathbb R_{\geq0}\)
be absolutely continuous,
\(
|\dot y(\tau)|
\leq L_{\dot y}
\),
and
\(
\int_{t-T}^{t}
y(\tau)d\tau
\leq\bar I
\),
where
\(L_{\dot y},\bar I,T>0\).

For \(0<\gamma_1<1\),
\begin{equation}
\int_{t-T}^{t}
y^{\gamma_1}(\tau)d\tau
\geq
\rho_{\gamma_1,T}
\left(
\int_{t-T}^{t}
y(\tau)d\tau
\right)^\mu,
\label{eq:low_integral_bound}
\end{equation}
where
\begin{equation}
\begin{aligned}
\rho_{\gamma_1,T}
:=
\min\Big\{
&
2^{\gamma_1-1}
L_{\dot y}^{-\frac{1-\gamma_1}{2}},
4^{\gamma_1-1}
T^{1-\gamma_1}
\bar I^{\frac{\gamma_1-1}{2}}
\Big\}.
\end{aligned}
\label{eq:rho_low}
\end{equation}
\end{lemma}

\begin{proof}
See Appendix~\ref{app:proof_derivative_limited_window}.
\end{proof}

For
\(
y(\tau)
=
V_d(\bx(\tau))
\),
define
\begin{equation}
\begin{aligned}
\rho_{\gamma_1,T}
=
\min\Big\{
&
2^{\gamma_1-1}
L_{\dot V}^{-\frac{1-\gamma_1}{2}},
4^{\gamma_1-1}
T^{1-\gamma_1}
\bar I_x^{\frac{\gamma_1-1}{2}}
\Big\}.
\end{aligned}
\label{eq:rho_state_window}
\end{equation}

\begin{theorem}[Practical fixed-time convergence of the learned closed loop]
\label{thm:closed_loop}
Suppose that
Assumption~\ref{ass:plant}
holds and that
Lemma~\ref{lem:identification_error}
and
Lemma~\ref{lem:low_window}
apply on a compact forward-invariant
operating domain.
Suppose further that the informative
replay stack is available throughout the critic-learning process,
the critic satisfies the nonzero-residual
conclusion of
Theorem~\ref{thm:fixed_time_critic},
and the preceding bounds on the policy
mismatch hold.

Let
\begin{equation}
\begin{aligned}
T_W
&:=
T_{W,\Delta},
\qquad
t_0:=T_W+T,
\\
\bar W_c
&:=
\sqrt{
2\lambda_{\max}(\bm\Gamma)
\bar V_W
},
\\
a_x
&:=
k_1\rho_{\gamma_1,T},
\qquad
b_x:=
k_2T^{1-\nu},
\\
\Delta_I
&:=
T\left[
b_1\bar W_c^{\gamma_1+1}
+
b_2\bar W_c^{\gamma_2+1}
+
\Delta_x
\right],
\\
h_1
&:=
\alpha_Wa_{W\Delta},
\qquad
h_2:=
\alpha_Wb_{W\Delta},
\\
C_1
&:=
\min\{a_x,h_1\},
\\
C_2
&:=
2^{1-\nu}
\min\{b_x,h_2\},
\\
\Delta_J
&:=
\Delta_I+\Delta_W.
\end{aligned}
\label{eq:closed_loop_constants}
\end{equation}

Assume
\(C_1,C_2>0\),
and select any
\(\theta_J\in(0,1)\).
Define
\(\bar{\mathcal J}\geq0\)
as the unique nonnegative solution of
\begin{equation}
(1-\theta_J)
\left[
C_1\bar{\mathcal J}^{\mu}
+
C_2\bar{\mathcal J}^{\nu}
\right]
=
\Delta_J,
\label{eq:closed_loop_residual_radius}
\end{equation}
with
\(\bar{\mathcal J}=0\)
when
\(\Delta_J=0\).

Then the set
\begin{equation}
\mathcal B_J
:=
\left\{
\mathcal J\geq0:
\mathcal J\leq\bar{\mathcal J}
\right\}
\label{eq:closed_loop_invariant_set}
\end{equation}
is forward invariant under the post-learning
closed-loop inequality, and
\(\mathcal J(t)\)
enters \(\mathcal B_J\) no later than
\begin{equation}
\begin{aligned}
T_{\mathrm{FT}}
&:=
t_0+T_X^{\mathrm{FT}},
\\
T_X^{\mathrm{FT}}
&:=
\frac{1}{
\theta_JC_1(1-\mu)}
+
\frac{1}{
\theta_JC_2(\nu-1)}.
\end{aligned}
\label{eq:closed_loop_fixed_time_bound}
\end{equation}
independently of the initial value of
\(\mathcal J(t_0)\).

Consequently, for
\(t\geq T_{\mathrm{FT}}\),
\begin{equation}
\begin{aligned}
V_W(t)
&\leq
\bar{\mathcal J},\quad
\|\tilde{\bW}(t)\|
\leq
\sqrt{
2\lambda_{\max}(\bm\Gamma)
\bar{\mathcal J}
},\quad
\int_{t-T}^{t}
V_d(\bx(\tau))d\tau
\leq
\bar{\mathcal J}.
\end{aligned}
\label{eq:closed_loop_practical_conclusions}
\end{equation}

\end{theorem}

\begin{proof}
From
Theorem~\ref{thm:fixed_time_critic},
the critic weight error satisfies
\(
\|\tilde{\bW}(t)\|
\leq
\bar W_c
\)
for all
\(t\geq T_W\).

Therefore, for
\(t\geq t_0=T_W+T\),
every point of the moving window
belongs to the post-learning interval,
and
\begin{equation}
\begin{aligned}
\int_{t-T}^{t}
\|\tilde{\bW}(\tau)\|^{\gamma_i+1}d\tau
\leq
T\bar W_c^{\gamma_i+1},
\quad i=1,2.
\end{aligned}
\label{eq:closed_loop_window_weight_bound}
\end{equation}

Moreover,
\(
\dot{\mathcal I}_x
=
V_d(\bx(t))
-
V_d(\bx(t-T))
=
\int_{t-T}^{t}
\dot V_d(\bx(\tau))d\tau
\).

Integrating
\eqref{eq:Vstar_learned_bound}
over
\([t-T,t]\)
gives
\begin{equation}
\begin{aligned}
\dot{\mathcal I}_x
\leq{}&
-k_1
\int_{t-T}^{t}
(V_d)^{\gamma_1}d\tau-
k_2
\int_{t-T}^{t}
(V_d)^\nu d\tau
\\
&+
b_1
\int_{t-T}^{t}
\|\tilde{\bW}\|^{\gamma_1+1}d\tau
+
b_2
\int_{t-T}^{t}
\|\tilde{\bW}\|^{\gamma_2+1}d\tau
+
T\Delta_x.
\end{aligned}
\label{eq:closed_loop_window_derivative}
\end{equation}

Lemma~\ref{lem:low_window}
gives
\(
\int_{t-T}^{t}
(V_d)^{\gamma_1}d\tau
\geq
\rho_{\gamma_1,T}
\mathcal I_x^\mu
\),
whereas Jensen's inequality for
\(\nu>1\)
yields
\(
\int_{t-T}^{t}
(V_d)^\nu d\tau
\geq
T^{1-\nu}
\mathcal I_x^\nu
\).

Using
\eqref{eq:closed_loop_window_weight_bound}
gives
\begin{equation}
\dot{\mathcal I}_x
\leq
-a_x\mathcal I_x^\mu
-b_x\mathcal I_x^\nu
+\Delta_I,
\qquad
t\geq t_0.
\label{eq:Ix_comparison}
\end{equation}

For the critic, the Lyapunov inequality
established in the proof of
Theorem~\ref{thm:fixed_time_critic}
is
\begin{equation}
\dot V_W
\leq
-h_1V_W^\mu
-h_2V_W^\nu
+\Delta_W,
\qquad
t\geq t_0.
\label{eq:VW_closed_loop_recall}
\end{equation}

Combining
\eqref{eq:Ix_comparison}
and
\eqref{eq:VW_closed_loop_recall}
gives
\begin{equation}
\begin{aligned}
\dot{\mathcal J}
\leq{}&
-a_x\mathcal I_x^\mu
-b_x\mathcal I_x^\nu
-
h_1V_W^\mu
-h_2V_W^\nu
+
\Delta_I+\Delta_W.
\end{aligned}
\label{eq:J_comparison_uncombined}
\end{equation}

Using
\(
a^\mu+b^\mu
\geq
(a+b)^\mu
\)
for
\(0<\mu<1\),
and
\(
a^\nu+b^\nu
\geq
2^{1-\nu}(a+b)^\nu
\)
for
\(\nu>1\),
one obtains
\begin{equation}
\dot{\mathcal J}
\leq
-C_1\mathcal J^\mu
-C_2\mathcal J^\nu
+\Delta_J,
\qquad
t\geq t_0.
\label{eq:J_comparison}
\end{equation}

For
\(\mathcal J>\bar{\mathcal J}\),
the monotonicity of the power functions
and
\eqref{eq:closed_loop_residual_radius}
imply
\(
\Delta_J
<
(1-\theta_J)
[
C_1\mathcal J^\mu
+
C_2\mathcal J^\nu
]
\).

Therefore, outside
\(\mathcal B_J\),
\begin{equation}
\dot{\mathcal J}
\leq
-\theta_JC_1\mathcal J^\mu
-\theta_JC_2\mathcal J^\nu.
\label{eq:J_outside_set}
\end{equation}

The fixed-time comparison argument,
applied until the first entrance into
\(\mathcal B_J\), gives
\begin{equation}
\begin{aligned}
T_{\mathcal B_J}-t_0
\leq{}&
\frac{1}{
\theta_JC_1(1-\mu)}
+
\frac{1}{
\theta_JC_2(\nu-1)}
=
T_X^{\mathrm{FT}},
\end{aligned}
\label{eq:J_entrance_time}
\end{equation}
independently of
\(\mathcal J(t_0)\).

At
\(\mathcal J=\bar{\mathcal J}\),
\eqref{eq:J_comparison}
and
\eqref{eq:closed_loop_residual_radius}
yield
\begin{equation}
\begin{aligned}
\dot{\mathcal J}
&\leq
-C_1\bar{\mathcal J}^\mu
-C_2\bar{\mathcal J}^\nu
+\Delta_J
=
-\theta_J
[
C_1\bar{\mathcal J}^\mu
+
C_2\bar{\mathcal J}^\nu
]
\leq0.
\end{aligned}
\label{eq:J_boundary_invariance}
\end{equation}

Thus, \(\mathcal B_J\) is forward
invariant, and
\(
\mathcal J(t)\leq\bar{\mathcal J}
\)
for all
\(t\geq T_{\mathrm{FT}}\),
where
\(
T_{\mathrm{FT}}
=
t_0+T_X^{\mathrm{FT}}
\).

Since
\(
\mathcal J(t)
=
\mathcal I_x(t)+V_W(t)
\)
and both terms are nonnegative,
it follows that
\(
V_W(t)\leq\bar{\mathcal J}
\)
and
\(
\mathcal I_x(t)\leq\bar{\mathcal J}
\)
for all
\(t\geq T_{\mathrm{FT}}\).

The critic-weight estimate follows from
\(
V_W
\geq
\|\tilde{\bW}\|^2/
[
2\lambda_{\max}(\bm\Gamma)
]
\),
which yields
\(
\|\tilde{\bW}(t)\|
\leq
\sqrt{
2\lambda_{\max}(\bm\Gamma)
\bar{\mathcal J}
}
\).

Moreover, by the definition of
\(\mathcal I_x(t)\),
\(
\int_{t-T}^{t}
V_d(\bx(\tau))d\tau
\leq
\bar{\mathcal J}
\)
for all
\(t\geq T_{\mathrm{FT}}\).

This completes the proof.
\end{proof}

\section{Simulation}
\subsection{Simulation setup}

All simulations are implemented in MATLAB R2026a.
The plant is integrated using a fourth-order Runge--Kutta
scheme with \(\Delta t=0.01\,\mathrm{s}\),
while the critic weights are updated online using
the proposed two-power learning law.
Consider the nonlinear control-affine plant
\(
\dot{\bx}
=
\bfv(\bx)+\bg(\bx)\bu
\),
where
\(
\bx=[x_1,x_2,x_3]^\top\in\mathbb R^3
\),
\(
\bu=[u_1,u_2,u_3]^\top\in\mathbb R^3
\),
and
\(
\bg(\bx)=\bm I_3
\).
The unknown drift is
\(
\bfv(\bx)
=
[f_1(\bx),f_2(\bx),f_3(\bx)]^\top
\),
with
\(
f_1(\bx)
=
-\tanh(x_1+x_2)-0.4x_1^3
\),
\(
f_2(\bx)
=
-\sin(x_2+x_3)-0.3x_1x_2
\),
and
\(
f_3(\bx)
=
-\frac{x_3}{1+x_3^2}
-0.2\tanh(x_1x_3)
\).
For drift identification, the dictionary is chosen as
\(
\bsigma_f(\bx)
=
[
\tanh(x_1+x_2),
x_1^3,
\sin(x_2+x_3),
x_1x_2,
\frac{x_3}{1+x_3^2},
\tanh(x_1x_3)
]^\top
\),
and the identified drift is represented by
\(
\hat{\bfv}(\bx)
=
\hat{\bW}_f^\top\bsigma_f(\bx)
\),
where
\(
\hat{\bW}_f\in\mathbb R^{6\times3}
\).
The desired value function is selected as
\(
V_d(\bx)
=
\frac12\bx^\top\bx
=
\frac12(x_1^2+x_2^2+x_3^2)
\),
with
\(
r^2(\bx)=\bx^\top\bx
\).
The critic basis functions are
\(
\bphi(\bx)
=
[
x_1^2,x_2^2,x_3^2,
x_1x_2,x_1x_3,x_2x_3,
x_1^4,x_2^4,x_3^4
]^\top
\),
and the ideal critic weights are
\(
\bW^*
=
[0.5,0.5,0.5,0,0,0,0,0,0]^\top
\).
For the inverse-optimal implementation,
define
\(
\hat F_V(\bx)
=
\nabla V_d^\top(\bx)\hat{\bfv}(\bx)
\)
and
\(
\hat A(\bx)
=
\hat F_V(\bx)
+
k_1V_d^{\gamma_1}(\bx)
+
k_2V_d^\nu(\bx)
\),
where
\(
k_1=0.8
\),
\(
k_2=0.6
\),
\(
\gamma_1=0.75
\),
\(
\gamma_2=2
\),
and
\(
\nu=(\gamma_2+1)/2=1.5
\).
For
\(
\bx\neq\bm0
\),
the associated inverse-optimal quantities are
\(
\bm L_1(\bx)
=
(2\hat A(\bx)/r^2(\bx)-1)\bx^\top
\),
\(
L_2(\bx)
=
-\hat F_V(\bx)
+
\hat A^2(\bx)/r^2(\bx)
\),
and
\(
\bu_c(\bx)
=
-\hat A(\bx)\bx/r^2(\bx)
\),
with the continuous extension
\(
\bu_c(\bm0)=\bm0
\)
and the control-weighting matrix
\(
\bR=\bm I_3
\).
The critic-learning parameters are chosen as
\(
\eta_{W1}=1.8
\),
\(
\eta_{W2}=0.45
\),
and
\(
\alpha_W=0.24
\),
with
\(
\bm\Gamma=\bm I_9
\).
Five simulations are performed over
\(
t\in[0,100]\,\mathrm{s}
\),
with initial state norms
\(
\|\bx(0)\|_2
\in\{0.6,1.0,1.4,1.8,2.2\}
\)
and different initial critic weights.
The same probing realization is employed
in all five cases to ensure a consistent
comparison of the closed-loop responses.


The five simulation cases use
\(
\|\bx(0)\|_2
\in\{0.6,1.0,1.4,1.8,2.2\}
\),
respectively, with initial critic weights
\(
\hat{\bW}_0^{(1)}=\bm0_9
\),
\(
\hat{\bW}_0^{(2)}=0.15\bm1_9
\),
\(
\hat{\bW}_0^{(3)}=-0.20\bm1_9
\),
\(
\hat{\bW}_0^{(4)}
=0.30[1,-1,1,-1,1,-1,1,-1,1]^\top
\),
and
\(
\hat{\bW}_0^{(5)}=0.25\bm1_9
\).



\subsection{Simulation Results}
\label{sec:simulation_results}

\begin{figure}[H]
    \centering
    \scalebox{1}[0.7]{%
        \includegraphics[width=\columnwidth]{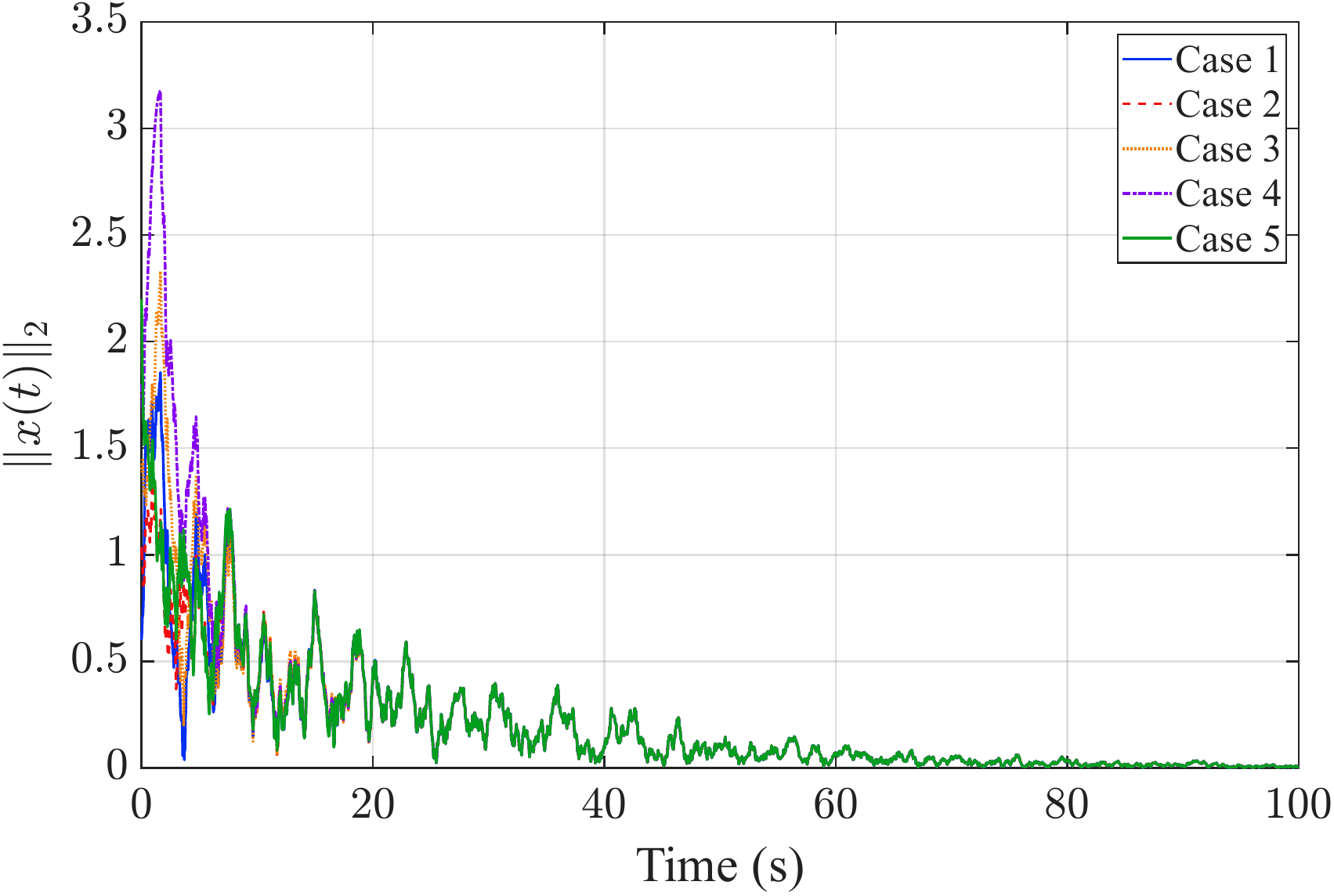}%
    }

\caption{Evolution of the state norm
\(\|\bx(t)\|_2\) under five different initial
conditions: Case 1,
\(\|\bx(0)\|_2=0.6\),
\(\hat{\bW}(0)=\bm0_9\);
Case 2,
\(\|\bx(0)\|_2=1.0\),
\(\hat{\bW}(0)=0.15\bm1_9\);
Case 3,
\(\|\bx(0)\|_2=1.4\),
\(\hat{\bW}(0)=-0.20\bm1_9\);
Case 4,
\(\|\bx(0)\|_2=1.8\),
\(\hat{\bW}(0)=
0.30[1,-1,1,-1,1,-1,1,-1,1]^\top\);
and Case 5,
\(\|\bx(0)\|_2=2.2\),
\(\hat{\bW}(0)=0.25\bm1_9\).}

    \label{fig:sim_state_norm}
\end{figure}

\begin{figure}[H]
    \centering
    \scalebox{1}[0.6]{%
        \includegraphics[width=\columnwidth]{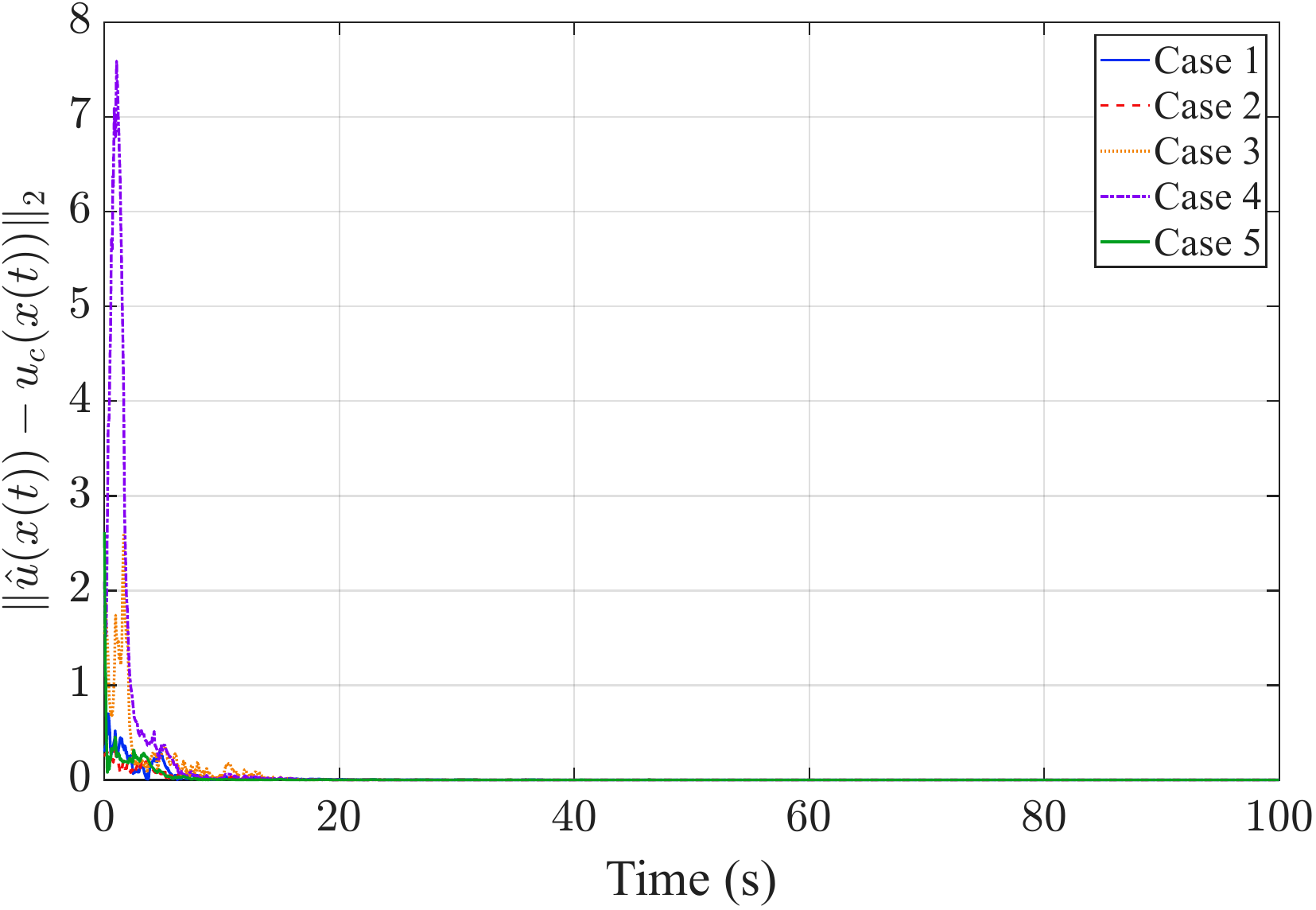}%
    }

\caption{Evolution of the learned-policy approximation error
\(\|\hat{\bu}(\bx(t))-\bu_c(\bx(t))\|_2\)
under five different initial conditions:
Case 1,
\(\|\bx(0)\|_2=0.6\),
\(\hat{\bW}(0)=\bm0_9\);
Case 2,
\(\|\bx(0)\|_2=1.0\),
\(\hat{\bW}(0)=0.15\bm1_9\);
Case 3,
\(\|\bx(0)\|_2=1.4\),
\(\hat{\bW}(0)=-0.20\bm1_9\);
Case 4,
\(\|\bx(0)\|_2=1.8\),
\(\hat{\bW}(0)=
0.30[1,-1,1,-1,1,-1,1,-1,1]^\top\);
and Case 5,
\(\|\bx(0)\|_2=2.2\),
\(\hat{\bW}(0)=0.25\bm1_9\).}

    \label{fig:sim_policy_error}
\end{figure}

\begin{figure}[H]
    \centering
    \scalebox{1}[0.6]{%
        \includegraphics[width=\columnwidth]{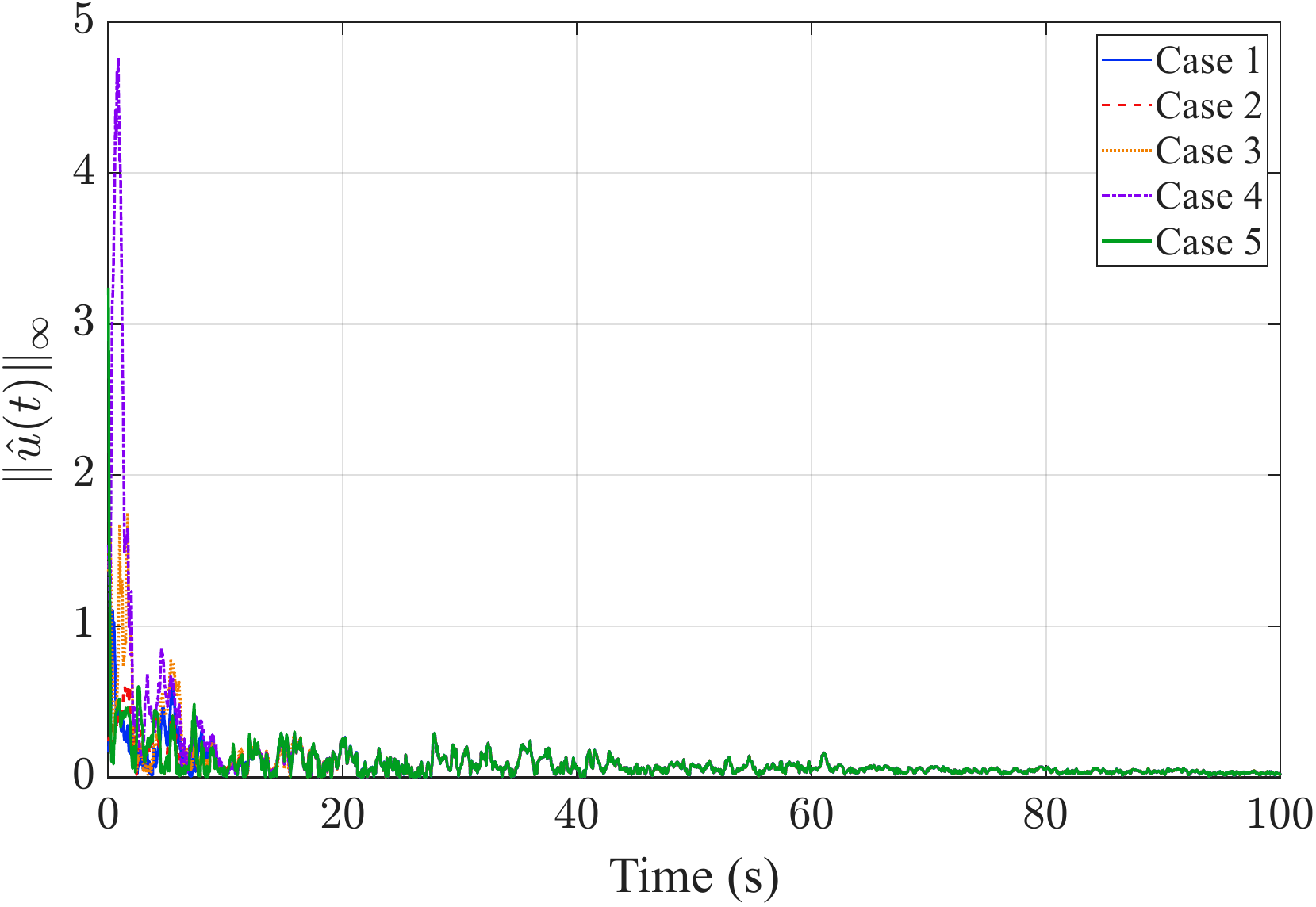}%
    }
    \caption{Evolution of the learned-control norm
    \(\|\hat{\bu}(t)\|_\infty\)
    under five different initial conditions:
    Case 1,
    \(\|\bx(0)\|_2=0.6\),
    \(\hat{\bW}(0)=\bm0_9\);
    Case 2,
    \(\|\bx(0)\|_2=1.0\),
    \(\hat{\bW}(0)=0.15\bm1_9\);
    Case 3,
    \(\|\bx(0)\|_2=1.4\),
    \(\hat{\bW}(0)=-0.20\bm1_9\);
    Case 4,
    \(\|\bx(0)\|_2=1.8\),
    \(\hat{\bW}(0)=
    0.30[1,-1,1,-1,1,-1,1,-1,1]^\top\);
    and Case 5,
    \(\|\bx(0)\|_2=2.2\),
    \(\hat{\bW}(0)=0.25\bm1_9\).}
    \label{fig:sim_control_norm}
\end{figure}

\begin{figure}[H]
    \centering
    \scalebox{1}[0.8]{%
        \includegraphics[width=\columnwidth]{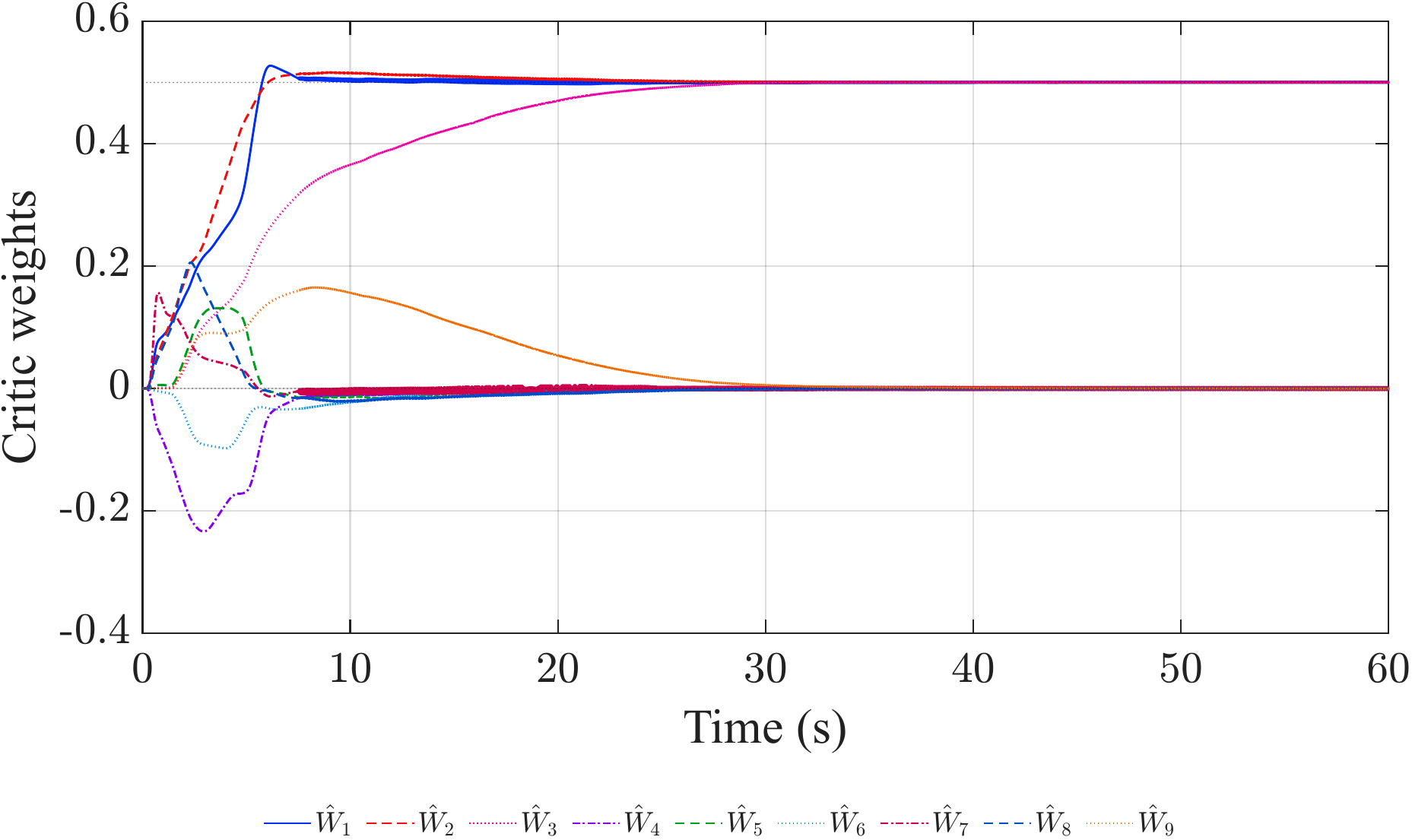}%
    }
    \caption{Evolution of the critic weights
    \(\hat{\bW}(t)\) during the online learning
    process for Case~1, corresponding to
    \(\|\bx(0)\|_2=0.6\) and
    \(\hat{\bW}(0)=\bm0_9\).}
    \label{fig:sim_critic_weights}
\end{figure}

\begin{figure}[H]
    \centering
    \scalebox{1}[0.8]{%
        \includegraphics[width=\columnwidth]{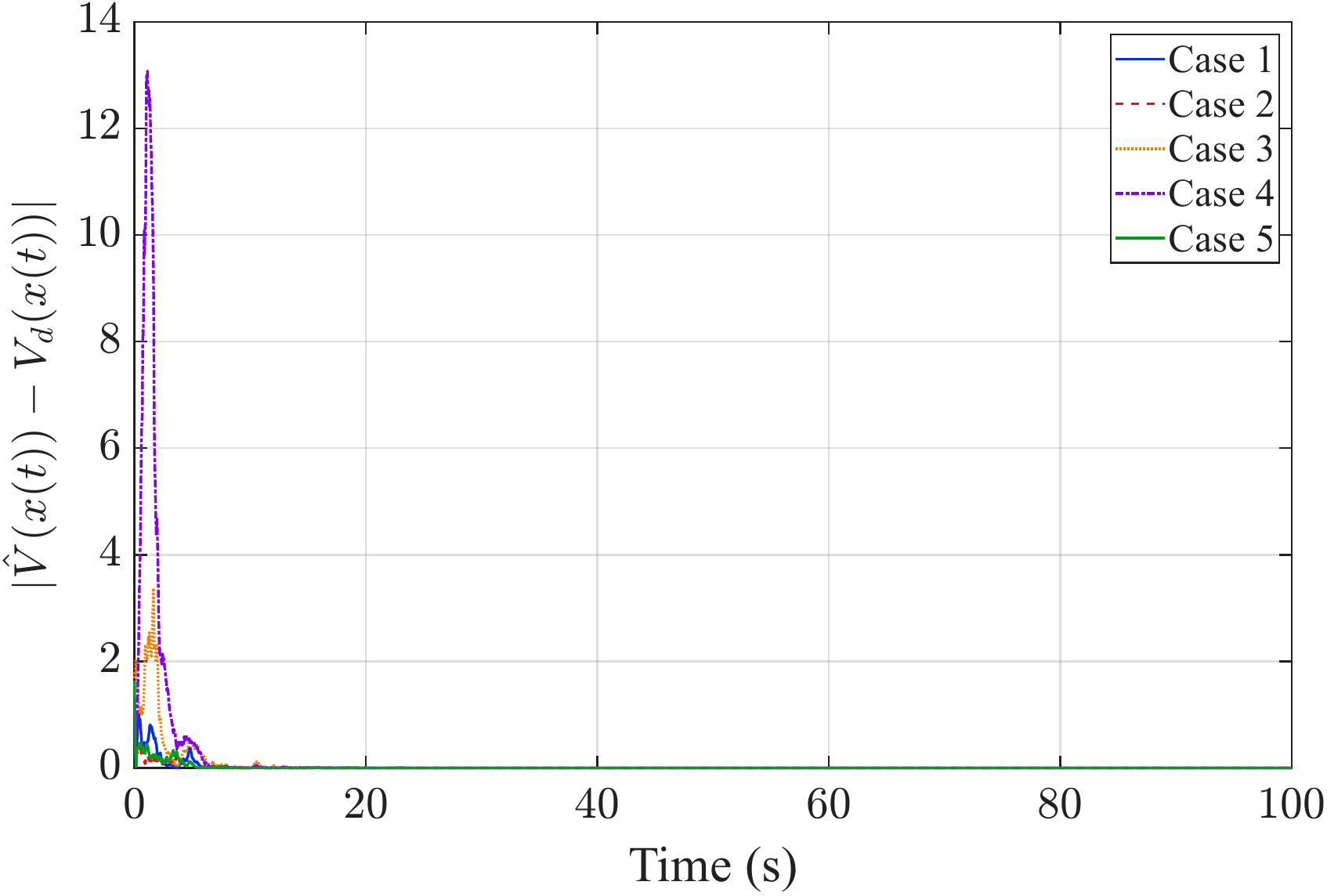}%
    }
    \caption{Evolution of the value-function approximation error
    \(|\hat{V}(\bx(t))-V_d(\bx(t))|\)
    under five different initial conditions.}
    \label{fig:sim_value_error}
\end{figure}

\begin{figure}[H]
    \centering
    \scalebox{1}[0.65]{%
        \includegraphics[width=\columnwidth]{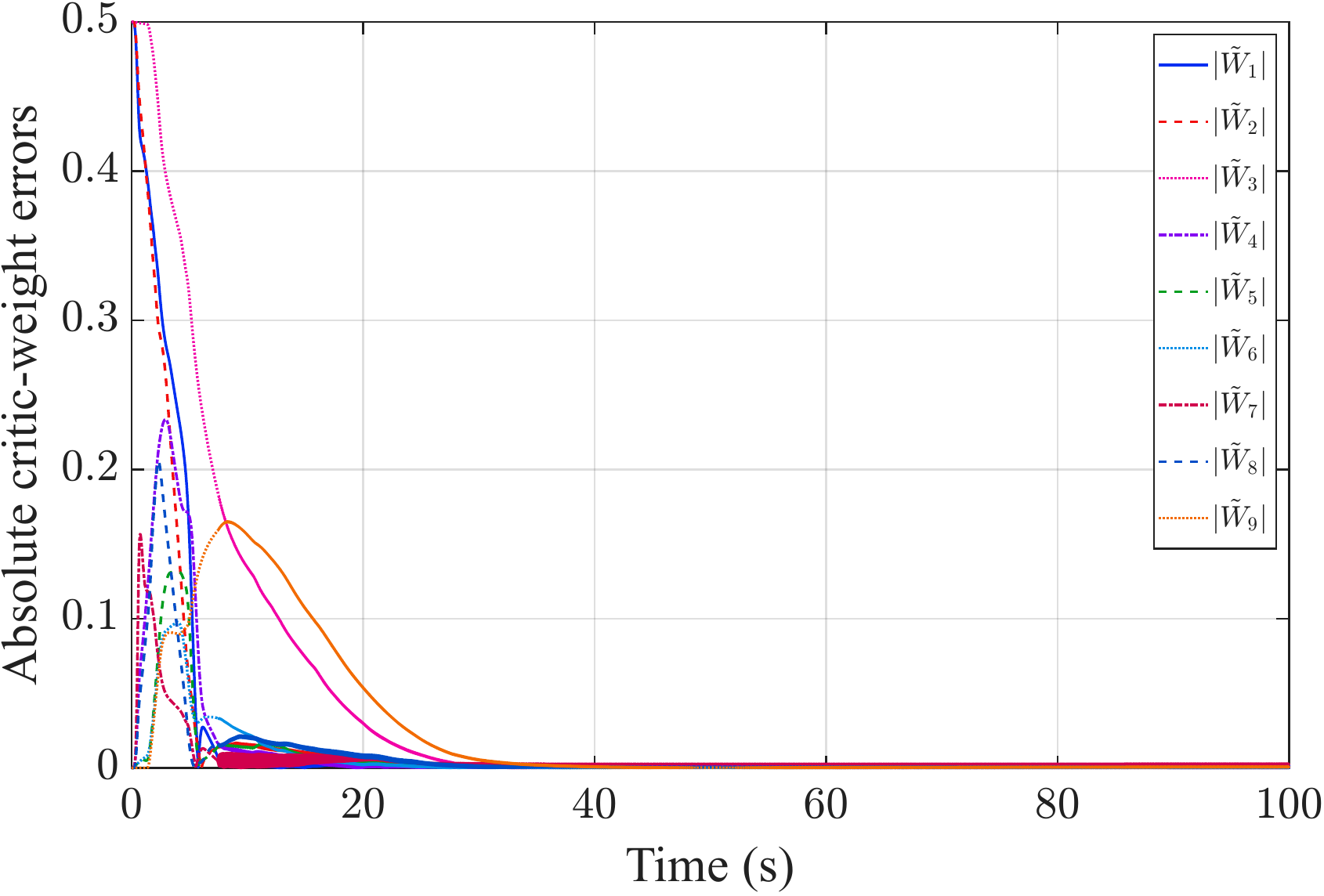}%
    }
    \caption{Evolution of the absolute critic-weight estimation errors
    for Case~1, corresponding to
    \(\|\bx(0)\|_2=0.6\) and
    \(\hat{\bW}(0)=\bm0_9\).}
    \label{fig:sim_weight_error_case1}
\end{figure}

\begin{figure}[H]
    \centering
    \scalebox{1}[0.65]{%
        \includegraphics[width=\columnwidth]{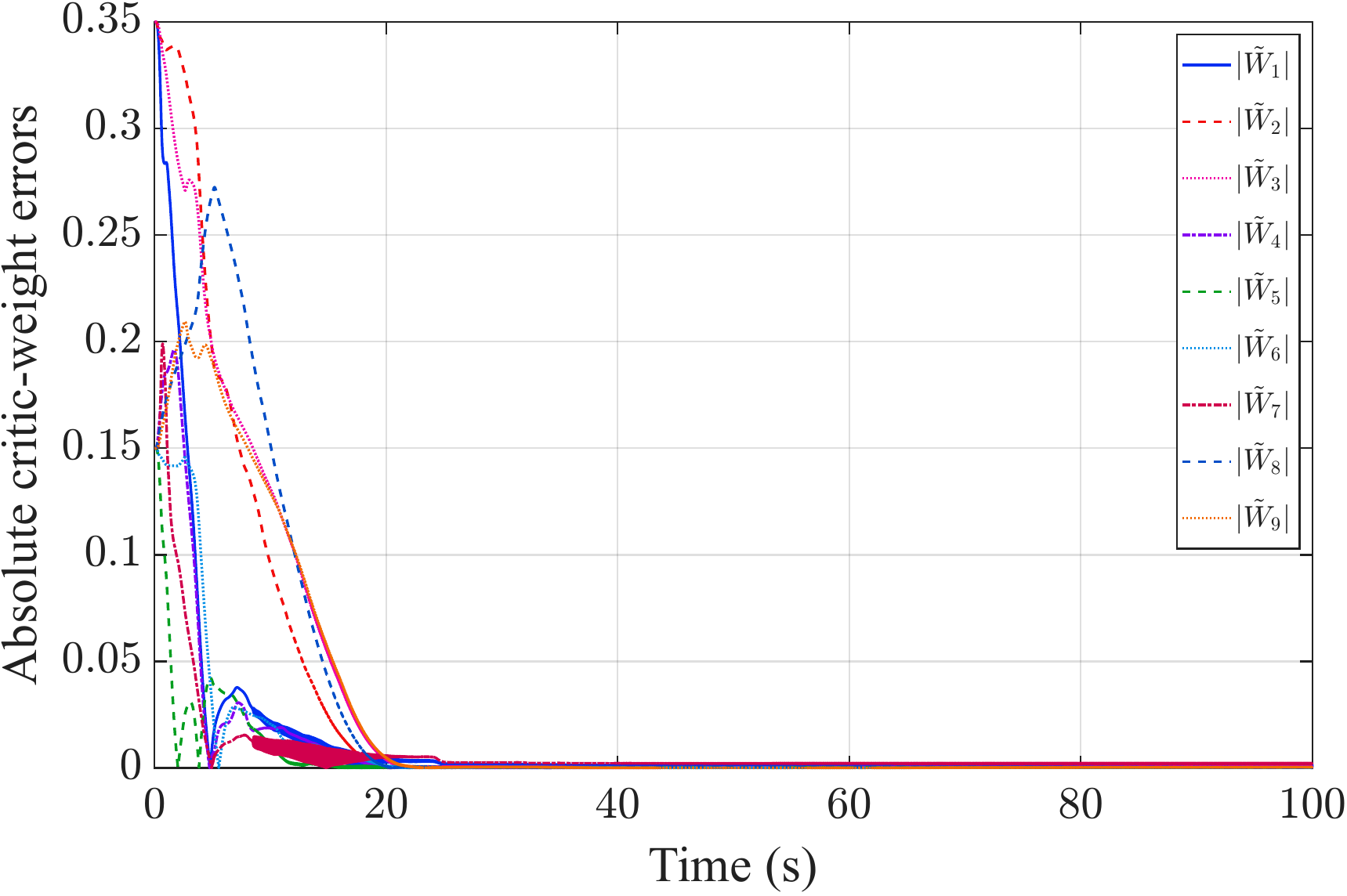}%
    }
    \caption{Evolution of the absolute critic-weight estimation errors
    for Case~2, corresponding to
    \(\|\bx(0)\|_2=1.0\) and
    \(\hat{\bW}(0)=0.15\bm1_9\).}
    \label{fig:sim_weight_error_case2}
\end{figure}

\begin{figure}[H]
    \centering
    \scalebox{1}[0.6]{%
        \includegraphics[width=\columnwidth]{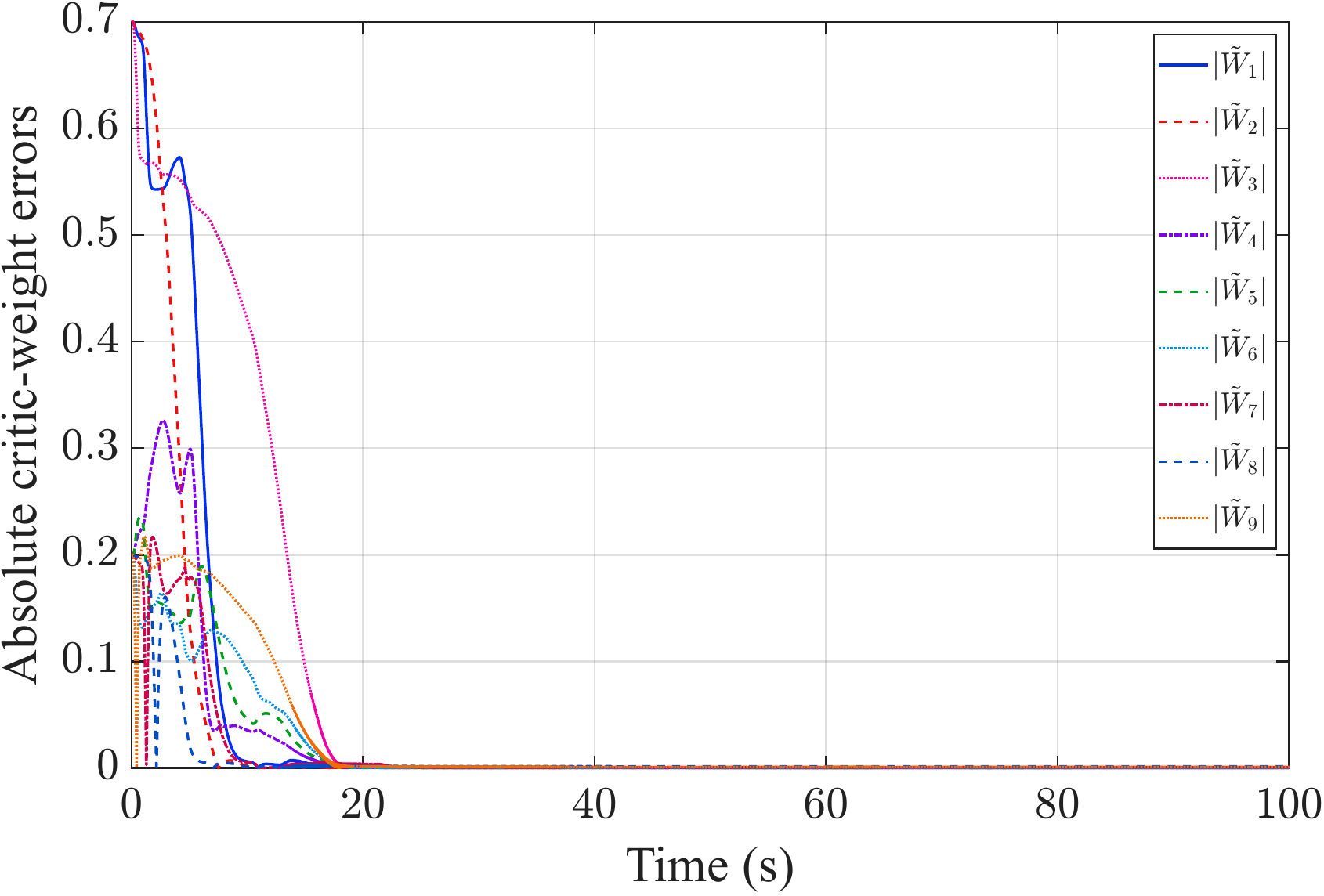}%
    }
    \caption{Evolution of the absolute critic-weight estimation errors
    for Case~3, corresponding to
    \(\|\bx(0)\|_2=1.4\) and
    \(\hat{\bW}(0)=-0.20\bm1_9\).}
    \label{fig:sim_weight_error_case3}
\end{figure}

\begin{figure}[H]
    \centering
    \scalebox{1}[0.65]{%
        \includegraphics[width=\columnwidth]{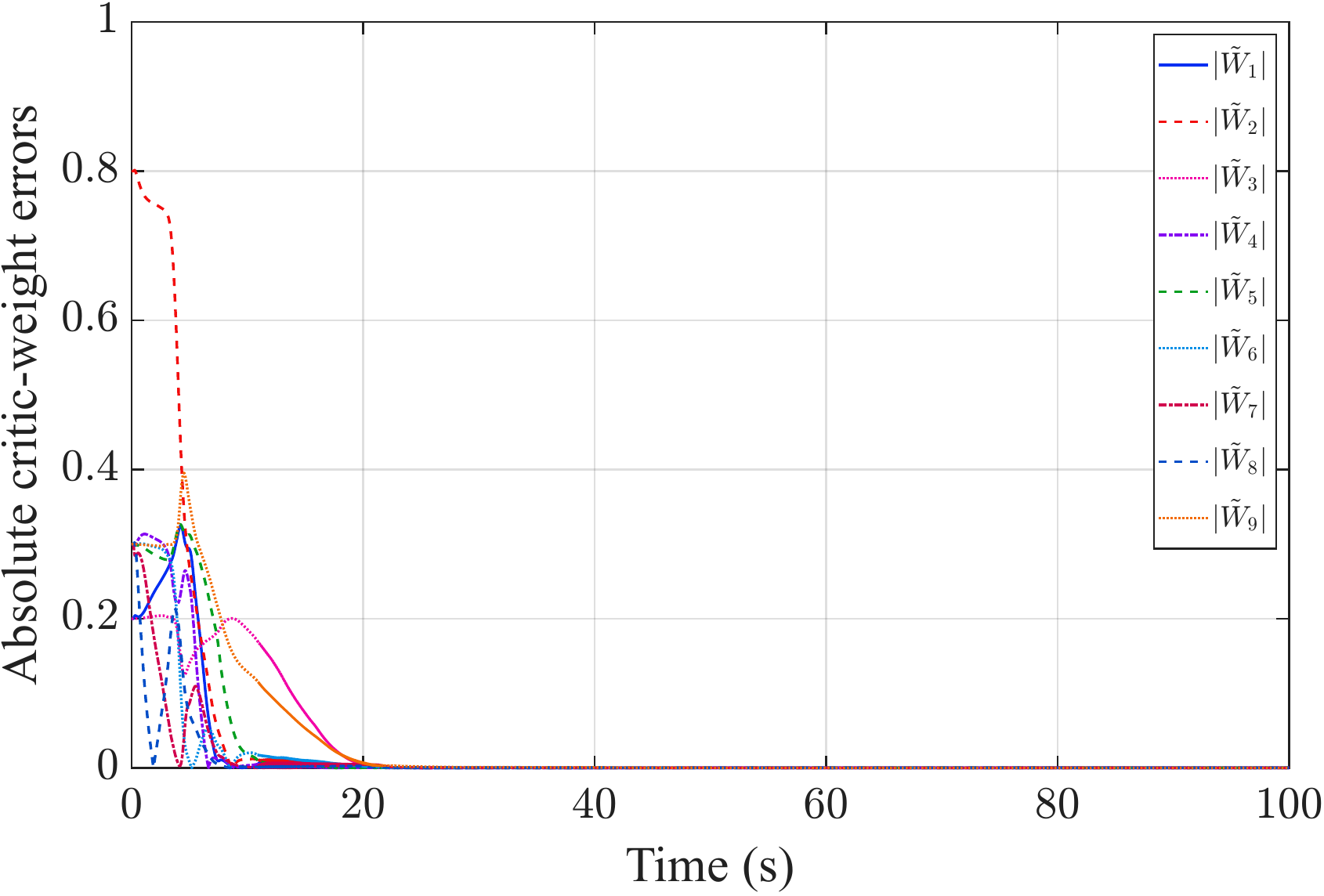}%
    }
    \caption{Evolution of the absolute critic-weight estimation errors
    for Case~4, corresponding to
    \(\|\bx(0)\|_2=1.8\) and
    \(\hat{\bW}(0)=
    0.30[1,-1,1,-1,1,-1,1,-1,1]^\top\).}
    \label{fig:sim_weight_error_case4}
\end{figure}

\begin{figure}[H]
    \centering
    \scalebox{1}[0.7]{%
        \includegraphics[width=\columnwidth]{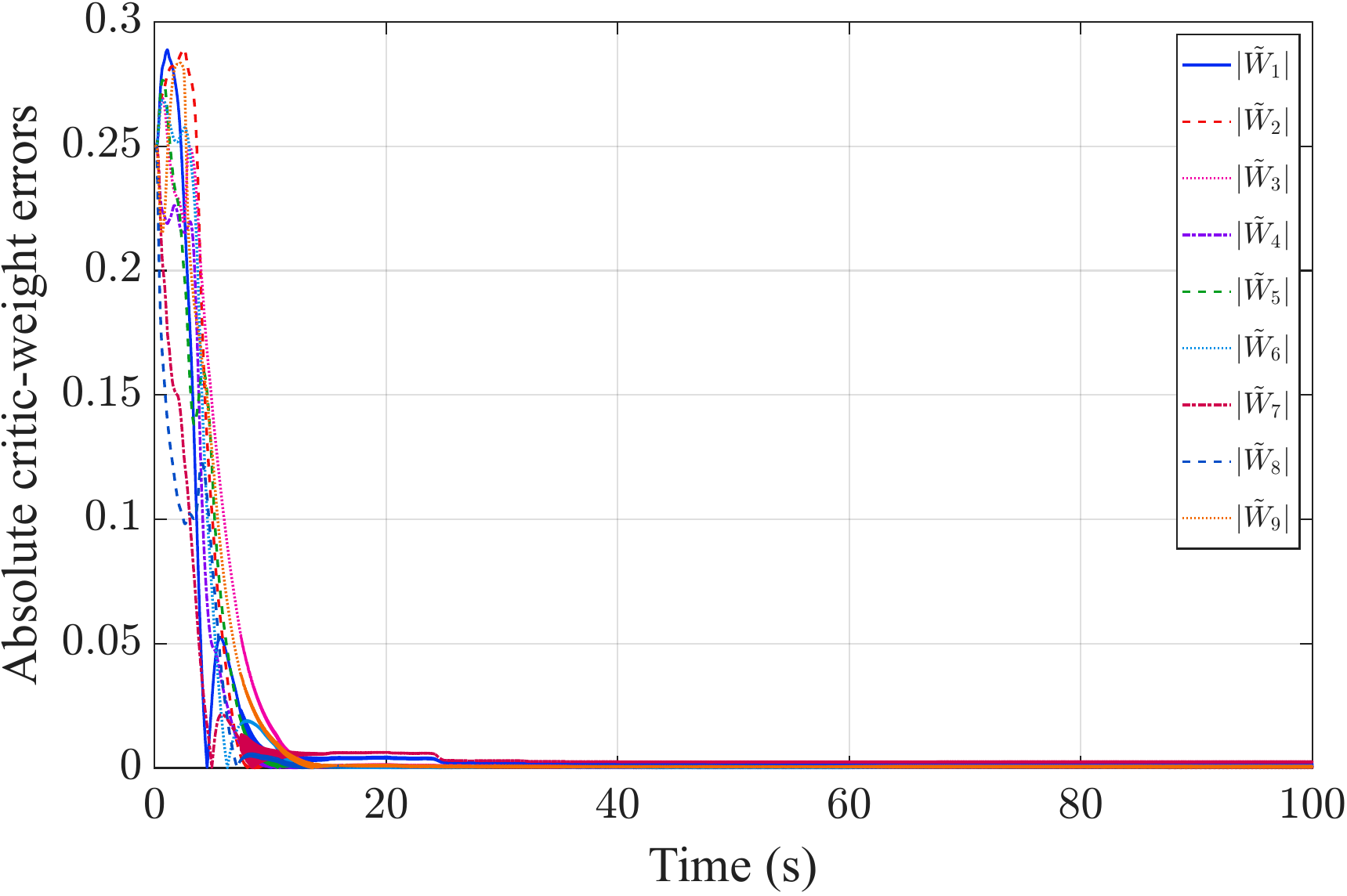}%
    }
    \caption{Evolution of the absolute critic-weight estimation errors
    for Case~5, corresponding to
    \(\|\bx(0)\|_2=2.2\) and
    \(\hat{\bW}(0)=0.25\bm1_9\).}
    \label{fig:sim_weight_error_case5}
\end{figure}

Figures~\ref{fig:sim_state_norm}--\ref{fig:sim_weight_error_case5}
illustrate the closed-loop and critic-learning performance
under five distinct initial conditions.
Figure~\ref{fig:sim_state_norm} shows that the state norms
decrease rapidly and subsequently approach a small
neighborhood of the origin, despite the different initial
state magnitudes and critic weights.
The transient oscillations observed during the early stage
are consistent with the simultaneous effects of online
learning and probing excitation.
Figures~\ref{fig:sim_policy_error} and
\ref{fig:sim_control_norm} show that the learned-policy
mismatch and control magnitude decrease substantially
after their initial transients, indicating that the learned
feedback progressively approaches the inverse-optimal
reference policy while maintaining bounded control
responses.
Figure~\ref{fig:sim_critic_weights} illustrates the
evolution of the critic parameters, whereas
Fig.~\ref{fig:sim_value_error} shows a rapid reduction
in the approximation error relative to the desired value
function.
More specifically,
Figs.~\ref{fig:sim_weight_error_case1}--
\ref{fig:sim_weight_error_case5} reveal nonmonotonic
transients in individual critic-weight errors, with the
largest visible errors initially reaching approximately
\(0.8\) in Case~4, while the errors in all five cases
subsequently approach small residual levels within
approximately \(20\)--\(30\,\mathrm{s}\).
These observations are consistent with the practical
fixed-time learning and closed-loop convergence
properties established theoretically, although the plots
alone do not verify the analytical upper bound
\(T_{\mathrm{FT}}\).
The relatively pronounced initial policy and
weight-estimation transients also indicate that convergence
quality depends on the interaction between excitation,
critic adaptation, and the initial learning mismatch.
Moreover, the use of an exactly representable desired
value function and a drift dictionary matched to the
simulated dynamics limits the assessment of robustness
against approximation and identification errors.
These limitations motivate future investigation of
excitation-aware adaptive learning gains and robust
fixed-time IRL under imperfect function approximation,
with explicit transient-control constraints and
quantitative residual-set verification.
Another promising direction is to extend the proposed
framework to safety-critical learning-based control,
where control barrier functions can be incorporated
to enforce state constraints throughout the learning
and control processes
\citep{Cohen2023SafetyCritical}.
Such an extension would require a unified stability
and safety analysis to establish forward invariance
of the admissible state set while preserving practical
fixed-time convergence guarantees under learning
and identification uncertainties.

\section{Conclusion}

\appendix

\section{Proof of Lemma~\ref{lem:low_window}}
\label{app:proof_derivative_limited_window}

\begin{proof}
Fix any \(t\) and define
\(I_y:=\int_{t-T}^{t}y(\tau)\,d\tau\).
If \(I_y=0\), then \(y(\tau)=0\) almost everywhere on
\([t-T,t]\), and the claimed inequality follows immediately.
Hence, suppose \(I_y>0\).

Since \(y\) is absolutely continuous on the compact interval
\([t-T,t]\), it is continuous and therefore attains its maximum.
Let
\(M_y:=\max_{\tau\in[t-T,t]}y(\tau)\),
and choose
\(\tau_m\in[t-T,t]\)
such that
\(y(\tau_m)=M_y\).
Because \(0<\gamma_1<1\), one has
\(y^{\gamma_1}(\tau)
=
y(\tau)y^{\gamma_1-1}(\tau)
\geq
M_y^{\gamma_1-1}y(\tau)\)
for every \(\tau\in[t-T,t]\).
Consequently,
\(\int_{t-T}^{t}y^{\gamma_1}(\tau)\,d\tau
\geq
M_y^{\gamma_1-1}I_y\).

We now distinguish two cases. First, suppose that
\(M_y\leq 4I_y/T\).
Then
\(M_y^{\gamma_1-1}
\geq
(4I_y/T)^{\gamma_1-1}\),
because \(\gamma_1-1<0\). Hence,
\(\int_{t-T}^{t}y^{\gamma_1}(\tau)\,d\tau
\geq
4^{\gamma_1-1}T^{1-\gamma_1}I_y^{\gamma_1}\).
Since
\(\mu=(\gamma_1+1)/2\),
one has
\(\gamma_1-\mu=(\gamma_1-1)/2<0\).
Using \(I_y\leq\bar I\) therefore gives
\(I_y^{\gamma_1-\mu}
\geq
\bar I^{(\gamma_1-1)/2}\).
It follows that
\(\int_{t-T}^{t}y^{\gamma_1}(\tau)\,d\tau
\geq
4^{\gamma_1-1}T^{1-\gamma_1}
\bar I^{(\gamma_1-1)/2}
I_y^\mu\).

Consider now the complementary case
\(M_y>4I_y/T\).
We first show that
\(M_y\leq L_{\dot y}T\).
Indeed, if \(M_y>L_{\dot y}T\), then the derivative bound
\(|\dot y(\tau)|\leq L_{\dot y}\)
implies
\(y(\tau)\geq
M_y-L_{\dot y}|\tau-\tau_m|\)
for all \(\tau\in[t-T,t]\).
Since
\(|\tau-\tau_m|\leq T\),
this gives
\(y(\tau)\geq M_y-L_{\dot y}T>0\).
More precisely, integrating the Lipschitz lower envelope over the
interval yields
\(I_y\geq M_yT-\frac12L_{\dot y}T^2\).
Under \(M_y>L_{\dot y}T\), the latter satisfies
\(I_y>M_yT/2\), which contradicts
\(M_y>4I_y/T\).
Thus,
\(M_y\leq L_{\dot y}T\).

At least one of the intervals
\([t-T,\tau_m]\) and \([\tau_m,t]\)
has length no smaller than \(T/2\).
Since
\(M_y/(2L_{\dot y})\leq T/2\),
there exists a one-sided subinterval adjacent to \(\tau_m\)
of length
\(M_y/(2L_{\dot y})\)
that is contained in \([t-T,t]\).
For every \(\tau\) in this subinterval, the derivative bound gives
\(y(\tau)\geq
M_y-L_{\dot y}|\tau-\tau_m|
\geq M_y/2\).
Therefore,
\(I_y\geq
(M_y/2)(M_y/(2L_{\dot y}))
=
M_y^2/(4L_{\dot y})\),
and hence
\(M_y\leq 2\sqrt{L_{\dot y}I_y}\).

Using again
\(\int_{t-T}^{t}y^{\gamma_1}(\tau)\,d\tau
\geq
M_y^{\gamma_1-1}I_y\)
and \(\gamma_1-1<0\), one obtains
\(\int_{t-T}^{t}y^{\gamma_1}(\tau)\,d\tau
\geq
(2\sqrt{L_{\dot y}I_y})^{\gamma_1-1}I_y\).
Since
\(\mu=(\gamma_1+1)/2\),
this reduces to
\(\int_{t-T}^{t}y^{\gamma_1}(\tau)\,d\tau
\geq
2^{\gamma_1-1}
L_{\dot y}^{-(1-\gamma_1)/2}
I_y^\mu\).

Thus, in either case,
\(\int_{t-T}^{t}y^{\gamma_1}(\tau)\,d\tau\)
is bounded below by
\(\rho_{\gamma_1,T}I_y^\mu\),
where \(\rho_{\gamma_1,T}\) is the coefficient defined in the
statement of the lemma. Substituting
\(I_y=\int_{t-T}^{t}y(\tau)\,d\tau\)
completes the proof.
\end{proof}

\section{Verification of Inverse Optimality with a Potentially Nonpositive Running Cost}
\label{app:running_cost_verification}

Following
\cite[Sec.~8.4, pp.~523--525]{Haddad2008NonlinearControl},
we verify the inverse-optimality property
of the running cost in~\eqref{eq:cost_function}. From
\eqref{eq:hamiltonian_predefined_time}--\eqref{eq:optimal_control_general} and Theorem~\ref{thm:inverse_optimal_pt}, the
Hamiltonian satisfies
\(H(\bx,\bu_c,\nabla V_d)=0\) and
\(\bm L_1^\top+\bg^\top\nabla V_d
=-2\bR\bu_c\). Hence,
\begin{equation}
\begin{aligned}
H(\bx,\bu,\nabla V_d)
&=H(\bx,\bu,\nabla V_d)
-H(\bx,\bu_c,\nabla V_d)\\
&=(\bu-\bu_c)^\top
\bR(\bu-\bu_c)\geq0.
\end{aligned}
\end{equation}
Equality holds uniquely at
\(\bu=\bu_c\).

From~\eqref{eq:cost_function}, completing the square gives
\begin{equation}
\begin{aligned}
\ell(\bx,\bu)
={}&\left(
\bu+\frac12\bR^{-1}\bm L_1^\top
\right)^\top\bR
\left(
\bu+\frac12\bR^{-1}\bm L_1^\top
\right)+L_2-\frac14
\bm L_1\bR^{-1}\bm L_1^\top.
\end{aligned}
\end{equation}
Thus, \(\ell(\bx,\bu)\) is not
necessarily pointwise nonnegative.


By Definition~\ref{def:admissible_secure_policy}, 
\(\Psi(\Omega)\) denote the class of
admissible policies, and assume that
\(\bu_c\in\Psi(\Omega)\).
For any \(\bu\in\Psi(\Omega)\),
the corresponding closed-loop state
satisfies \(\lim_{t\to\infty}\bx(t)=\bm0\).
Since \(V_d\) is continuous at the origin
and \(V_d(\bm0)=0\), it follows that
\(\lim_{T\to\infty}V_d(\bx(T))=0\).
Using \eqref{eq:cost_function} and
\eqref{eq:hamiltonian_predefined_time},
\begin{equation}
\begin{aligned}
J(\bx_0,\bu)
&=\lim_{T\to\infty}
\int_0^T
[H(\bx,\bu,\nabla V_d)-\dot V_d]\,dt\\
&=V_d(\bx_0)
-\lim_{T\to\infty}V_d(\bx(T))\\
&\quad+\int_0^\infty
H(\bx,\bu,\nabla V_d)\,dt\\
&=V_d(\bx_0)
+\int_0^\infty
(\bu-\bu_c)^\top\bR
(\bu-\bu_c)\,dt\\
&\geq V_d(\bx_0).
\end{aligned}
\end{equation}
For \(\bu=\bu_c\),
\(H(\bx,\bu_c,\nabla V_d)=0\);
therefore,
\begin{equation}
\begin{aligned}
J(\bx_0,\bu_c)
&=V_d(\bx_0)=\min_{\bu\in\Psi(\Omega)}
J(\bx_0,\bu).
\end{aligned}
\end{equation}
Moreover, using the prescribed
fixed-time decay
\eqref{eq:desired_fixed_time_decay},
\begin{equation}
\begin{aligned}
\ell(\bx,\bu_c)
&=-\dot V_d(\bx)=k_1V_d^{\gamma_1}(\bx)
+k_2V_d^{\frac{\gamma_2+1}{2}}(\bx)>0,
\qquad \bx\neq\bm0.
\end{aligned}
\end{equation}
This completes the proof.


\bibliographystyle{elsarticle-harv}
\bibliography{references}





\end{document}